\pdfoutput=1
\documentclass[traditabstract]{aa}
\usepackage{graphicx,psfrag,amsmath,color}

\usepackage{amsmath} 
 
\usepackage{graphicx}
\usepackage{txfonts}
\usepackage{natbib} 
\bibpunct{(}{)}{;}{a}{}{,} 

\begin{document} 

\title{Weighing the local dark matter with RAVE red clump stars}

\titlerunning{Weighing the local dark matter}

\author{O. Bienaym\'e\inst{1}, B. Famaey\inst{1}, A. Siebert\inst{1}
\and K.~C.~ Freeman\inst{2}
\and B.~K.~ Gibson\inst{3,4}
\and G.~Gilmore\inst{5}
\and E.~K.~ Grebel\inst{6}
\and J.~ Bland-Hawthorn\inst{7}
\and G.~Kordopatis\inst{5}
\and U. Munari\inst{8}
\and J.~F.~ Navarro\inst{9}
\and Q.~Parker\inst{10,11}
\and W. Reid\inst{10,11}
\and G.~M.~ Seabroke\inst{12}
\and A. Siviero\inst{13}
\and M.~Steinmetz\inst{14}
\and F.~Watson\inst{15}
\and R.~F.ÊG. Wyse\inst{16}
\and T.~Zwitter\inst{17}
}

\institute{Observatoire astronomique de Strasbourg, Universit\'e de Strasbourg, CNRS, UMR 7550, 11 rue de l'Universit\'e, F-67000 Strasbourg, France 
	\and
Mount Stromlo Observatory, RSAA, Australian National University, Weston Creek, Canberra, ACT 2611, Australia
	\and
Institute for Computational Astrophysics, Dept of Astronomy \& Physics, Saint MaryÕs University, Halifax, NS, BH3 3C3, Canada
	\and
Jeremiah Horrocks Institute, University of Central Lancashire, Preston, PR1 2HE, United Kingdom	
	\and
Institute of Astronomy, Cambridge University, Madingley Road, Cambridge CB3 0HA, UK
	\and
Astronomisches Rechen-Institut, Zentrum f\"ur Astronomie der
Universit\"at Heidelberg, M\"onchhofstr.\ 12--14, 69120 Heidelberg,
Germany 
	\and
Sydney Institute for Astronomy, School
of Physics A28, University of Sydney, NSW 2006
	\and
INAF Osservatorio Astronomico di Padova, 36012 Asiago (VI), Italy
	\and
Department of Physics and Astronomy. University of Victoria. Victoria, BC. Canada V8P 5C2
	\and
Department of Physics, Macquarie University, Sydney, NSW 2109, Australia 
	\and
Research Centre for Astronomy, Astrophysics and Astrophotonics, Macquarie University, Sydney, NSW 2109, Australia.  Australian Astronomical Observatory, PO Box 915, North Ryde, NSW 1670, Australia
	\and
Mullard Space Science Laboratory, University College London, Holmbury St Mary, Dorking, RH5 6NT, UK 
	\and
Department of Physics and Astronomy, Padova University, Vicolo dellÕOsservatorio 2, I-35122 Padova, Italy
	\and
Leibniz-Institut f\"ur Astrophysik Potsdam (AIP), An der Sternwarte 16, D-14482 Potsdam, Germany
	\and
Australian Astronomical Observatory, PO Box 915, North Ryde, NSW 1670, Australia
	\and
Johns Hopkins University, Homewood Campus, 3400 N Charles Street, Baltimore, MD 21218, USA	
	\and
Faculty of Mathematics and Physics, University of Ljubljana, 1000 Ljubljana, Slovenia
	 }


\date{Received / Accepted }

\abstract{We determine the Galactic potential in the solar neigbourhood from RAVE observations. We select red clump stars for which accurate distances, radial velocities, and metallicities have been measured. Combined with data from the 2MASS and UCAC catalogues, we build a sample of $\sim$4600 red clump stars within a cylinder of 500\,pc radius oriented in  the direction of the South Galactic Pole, in the range of 200\,pc to 2000\,pc distances. We deduce the vertical force and the total mass density distribution up to 2 kpc away from the Galactic plane by fitting a distribution function depending explicitly on three isolating integrals of the motion in a separable potential locally representing the Galactic one with four free parameters. Because of the deep extension of our sample, we can determine  nearly independently the dark matter mass density and the baryonic disc surface mass density. We find (i) at 1\,kpc $K_z/(2\pi G)=68.5\pm1.0\, $M$_{\sun}$\,pc$^{-2}$, and (ii) at 2\,kpc  $K_z/(2\pi G)=96.9\pm2.2\, $M$_{\sun} $pc$^{-2}$. Assuming the solar Galactic radius at $R_0=8.5$ kpc, we deduce the local dark matter density $\rho_{\rm DM}(z=0)=0.0143\pm0.0011$M$_{\sun}\,$pc$^{-3}=0.542\pm0.042\,{\rm Gev\, cm^{-3}}$ and the baryonic surface mass density $\Sigma_{\rm bar}=44.4\pm4.1\,$M$_{\sun}\,$pc$^{-2}$. Our results are in agreement with  previously published $K_z$ determinations up to 1\,kpc, while the extension to 2\,kpc shows some evidence for an unexpectedly large amount of dark matter. A flattening of the dark halo of order 0.8 can produce such a high local density in combination with a circular velocity of 240\,km\,s$^{-1}$.
 It could also be consistent with a spherical cored dark matter profile whose density does not drop sharply with radius.
 Another explanation, allowing for a lower circular velocity, could be the presence of a secondary dark component, a very thick disc  resulting either from the deposit of dark matter from the accretion of multiple small dwarf galaxies, or from the presence of an effective 'phantom' thick disc in the context of effective galactic-scale modifications of gravity.

}{}{}{}{}

\keywords{Galaxies: kinematics and dynamics }

\maketitle

\section{Introduction}

The complexity of the structure, dynamics, and history of our Galaxy is progressively unveiled with the recent advent of numerous  large surveys. The access to positions, velocities,  and chemical abundances with reasonable accuracy for large samples of stars allows us to explore  the detailed properties of our own Galaxy. In the near future, covering a huge Galactic volume with an unprecedented accuracy, Gaia observations will  revolutionize our understanding of Galactic formation and evolution \citep{per01}.

For the time being, we concentrate on the question of the dynamical estimate of the mass distribution in the solar neighbourhood, the $K_{\rm z}$ problem (where $K_{\rm z}$ is the vertical force perpendicular to the Galactic plane). We note that a major difficulty, in  any survey analysis, is the identification of systematic errors when they are an order of magnitude below the dispersions of the measurement errors. By selecting a sample of red clump (RC)  stars from the 
  RAVE survey \citep[RAdial Velocity Experiment survey,][]{kor13}, we can drastically  improve  the measurements needed for the $K_z$ problem  in terms of number or  precision. In particular, for the first time, we succeed in separating the force contributions from the Galactic discs and from the dark halo, and we find that the local density of dark matter is higher than previously expected.

Besides a better description of the local morphology of our Galaxy, our determination of a large local density of dark matter has implications on the morphology of the dark matter component(s) and also has implications for the terrestrial experiments of direct detection of the dark matter. The local density of dark matter can also provide a  test for the MOND effective theory \citep{Mil83,FamMc12}.
\\

Our new dynamical determination of the Galactic potential and $K_{\rm z}$ force perpendicular to the Galactic plane is  based on RAVE observations \citep[DR4,][]{kor13} towards the South Galactic Pole (SGP).
This paper is an extension of the previous works published by \citet{sou03}, 
\citet{sie03}, \citet{bie06}, and \citet{sou08}, which probed the properties of RC stars  within 100\,pc of the Sun and at larger distances towards the North Galactic Pole (NGP). 
 
The advantage of RC stars is that their distances are quite accurately  deduced from photometric measurements as long as such stars have been identified to belong to the red clump, usually from their colour and gravity \citep[see][for an early recognition of the red clump]{can69}.

From previous RAVE data releases  \citep{ste06,zwi08,sie11}, RC stars  were already used to probe the Galactic structure, for instance for the identification of stellar populations \citep{vel08}, for a first measurement of the velocity ellipsoid  tilt \citep{sie08}, or as a probe of the local 3D velocity field in a large volume of the solar neighbourhood \citep{Sie11a,wil13}.  

Here, we consider a   sample of about 4600 RC stars, mainly selected in the direction of the SGP up to distances of 2\,kpc from the Galactic plane.

A novelty of this work, setting aside the size of the  sample of stars with accurate distances, is our ability to measure the vertical potential further away from the plane, up to a vertical distance of 2\,kpc. From this, we deduce the local surface mass density and we  constrain  the shape of  the total vertical mass distribution. 

In this paper, the methods developed for the data analysis and modelling are the standard methods, except for the introduction of a separable St\"ackel potential with an explicit third integral of the motion to allow for the analysis at  $z$ distances higher than 1\,kpc. In addition, the selection function of RAVE observations of RC stars is well defined and accurately  determined. The completeness of observed RAVE RC stars towards the SGP is 83\% at 700\,pc,  66\% at 1.5\,kpc, and 20\% at 2\,kpc.\\

The history of the $K_{\rm z}$  measurements  covers decades of Galactic astronomy, and various summaries can be found in previous publications \citep{rea14}. We just mention here that a turning point was the advent of the Hipparcos satellite observations, which were decisive by increasing the amount of accurate data and critical by allowing us precise calibrations of the absolute magnitude of stars. In particular, this new set of data from the Hipparcos survey \citep{esa97} allowed us to map the kinematics and  variation of the stellar density  close to the plane, resulting in  a  redetermination of the Oort limit \citep{cre98,hol04}.  A detailed bibliography of the most recent works about the $K_z$ force, during  the last decade, can be found in  two recent publications \citep{gar12,zha13}, and in a  review on the $K_z$ problem and related questions by \citet{rea14}. \\
\\
This paper proceeds as follows. In Section 2, we present  the sample selection, the determination of the vertical volume density of the RC stars, and their kinematics.  Section 3 is devoted to the methods used to interpret the data. In Section 4, we present the results. Finally, conclusions are given and discussed in Sections 5 and 6.\\
 

\section{RC star selection, vertical density,  and kinematics}

\subsection{The sample selection}

\begin{figure}[!htbp]
\begin{center}
\resizebox{8.5cm}{!}{\rotatebox{-90}{\includegraphics{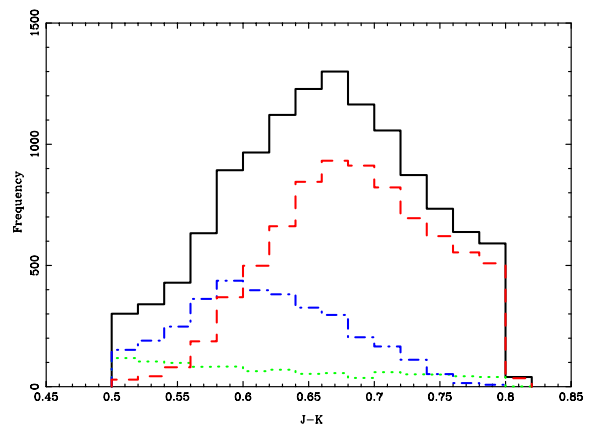}}}
\end{center}
  \caption{ Histograms of $J-K$ colours for 12308 RAVE stars observed towards the South Galactic Pole (black continuous line). Red clump stars (red dashed), subgiants (blue  dash-dotted line), dwarfs (green dotted line) identified from their gravity. 
  }
    \label{fig1}
\end{figure}

A first sample is drawn from the fourth data release of the RAVE survey \citep{kor13}  that contains stellar measurements of about 480\,000 stars. Measurements for each star include radial velocity, effective  temperature, gravity and metallicity. Cross-identification with the 2MASS and UCAC3 catalogues give us  the corresponding photometry and proper motions. Our aim is to select RC stars towards the SGP within a 500\,pc-radius cylinder centred at the solar position, and with its main axis directed towards the SGP. The use of a cylinder for counts, instead of using a cone, allows us to increase the number of stars at low $z$, and  to minimize the Malmquist-type bias. To minimize the consequence of interstellar extinction, we  also restrict the selection  to Galactic directions  $|b|>22\deg$. Thus, taking into consideration  the extinction $A_K$ = 0.0116 and a scale height for the extinction of 90\,pc \citep{gro08}, we  neglect  a  systematic overestimation of distances that ranges between 0.6 and 1.5\%. To include most of the RC stars and reject AGB stars, we apply as a colour selection $J-K$ within the range from 0.5 to 0.8,  based on 2MASS magnitudes (Figure \ref{fig1}).


\begin{figure}[!htbp]
\begin{center}
\resizebox{8.5cm}{!}{\rotatebox{-90}{\includegraphics{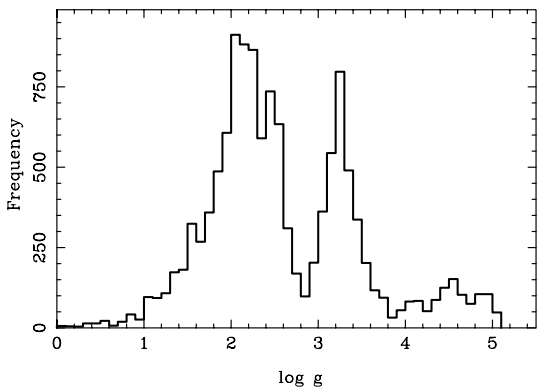}}}
\end{center}
  \caption{ Histogram of gravities  for 12308 RAVE stars observed towards the South Galactic Pole. 
  }
    \label{fig2}
\end{figure}

Based on  high-resolution spectroscopy of nearby Hipparcos stars,  gravity measurements   show that the red clump is defined by a restricted  range of gravities   from 1.8 to 2.8 \citep{sou08,val10}. This is in agreement with expectations from stellar evolution models \citep{gir00}, though they may suffer from uncertainties on the amount of mass loss   during the first ascent of the giant branch. The histogram of gravities from the RAVE DR4 stars with $J-K\in [0.5-0.8]$  shows a well-defined peak, allowing us to distinguish RC stars  from subgiants and dwarfs (Figure \ref{fig2}). We note a tail, containing a small fraction of stars with gravities smaller than 1.8, which might correspond to stars with lower absolute magnitudes along the red giant branch or along the AGB. Considering our colour selection interval, we do not expect such a significant number of these stars,  seen neither among nearby Hipparcos stars nor in model predictions. Here, we believe that their presence mainly results from  the uncertainties in the determination of RAVE gravities for giants. An indication for this  is   the small bias  in this range of gravities shown in  the figure 6 of \citet{kor13},  where  RAVE and external gravities are compared. For this reason, we choose the gravity  interval, 1.0 to 2.8, to define our RC sample. Using a sample by restricting  $\log g$ within  1.8 and 2.8, reduces the sample by 20\%. This does not modify our final conclusions concerning the measurement of the $K_z$ force. It is a complementary indication that these stars, with RAVE $\log g \in[1.,1.8]$ are also RC stars.

Finally, we only select stars with  a proper motion accuracy  better than 4 mas\,y$^{-1}$ and better than 5 km\,s$^{-1}$ for the radial velocity, with S/N greater than 20 in the  RAVE spectra, and $RV_{sky correction} <10$ . We  reject stars without measurement of gravity or metallicity. In the case of multiple observations of the same star (about 10 percent of all measured stars), we keep the measurements with the highest S/N.

\subsection{The RC stellar density}

The RC stars are located in the HR diagram  within restricted intervals of   colour, absolute magnitude, and  gravities.  Using the last Hipparcos reduction \citep{van07},   \citet{gro08} determined the mean absolute magnitude of RC stars, and found $M_K=-1.54$ with  a dispersion of 0.22 magnitude,  but 0.15 just considering the stars with the most accurately measured parallaxes. As noted by Groenevegen, his results could be affected by the photometry of bright stars that are saturated in the 2MASS data. 
This was confirmed by \citet{lan12} who measured new IR photometry of 226 RC  Hipparcos stars. Repeating the analysis, they obtained $M_K=-1.613\pm 0.015$. We note that the mean absolute $K$-magnitude of RC stars does not depend  on metallicity \citep{gro08}.

To define our main sample, we apply  the supplementary selection criterion:
\begin{equation}
m_K<-1.613+5 \log (500) -5 - 5 \log (\cos |b|).
\end{equation}
 This criterion includes RC stars located within the 500\,pc-radius cylindric volume oriented towards the Galactic poles. The  sample
 contains 9522 stars with $J-K$ within [0.5,0.8], among which there are 5618 RC stars.

A second sample, extracted from the 2MASS catalogue,  is used to determine the degree of completeness of our main RAVE sample and to quantify the selection function of RAVE observations. For this sample, the same criteria on $J-K$ colour, apparent magnitudes (Eq. 1), and Galactic directions are applied.

RAVE is a magnitude-limited survey of stars randomly selected in the southern celestial hemisphere. The original design was to only observe stars in the interval $9 < I < 12$, but the actual selection 
function includes stars both brighter and fainter.
For  sufficiently small intervals of magnitudes and Galactic directions, RAVE stars are randomly selected\footnote{At very low $b$ latitudes, not considered here, a colour cut criterion has been applied \citep[see][]{kor13}}. Hence,  for these intervals, the fraction of RC stars in the sky can be directly estimated from the ratio of the number of RAVE RC stars to the number of observed RAVE stars.

We determine this ratio using RAVE counts in both north and south Galactic directions in order to improve the statistics at low $z$. The 2MASS counts are   complete down to magnitude $K=14.3$ \citep{skr06}. Hence,  we combine 2MASS counts, RAVE counts, and RAVE RC counts to  estimate the total number of RC stars at any  interval of K magnitudes and directions where RAVE observations exist.

In practice,  since we cover a large range of Galactic latitudes and are interested in RC counts versus the vertical Galactic  distance $z$, we do not determine the counts using the apparent magnitudes but using  the corrected apparent magnitudes:
 
\begin{equation}
 K_C=m_K+ 5 \log (\sin |b|)=M_K+5\log |z|-5, 
\end{equation}
 
 which depends only on the absolute magnitude of the star and of its $|z|$ position.\\
 

\begin{figure}[!htbp]
\begin{center}
\resizebox{8.5cm}{!}{\rotatebox{-90}{\includegraphics{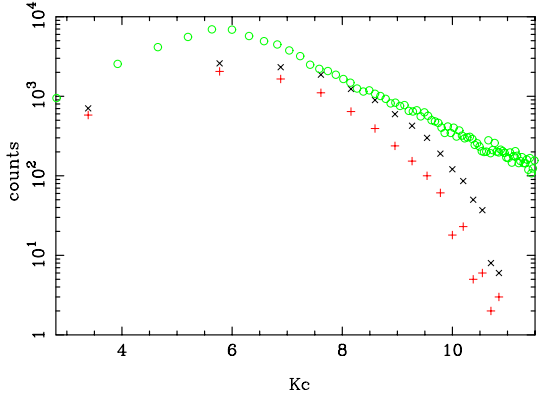}}}
\end{center}
  \caption{ 2MASS star counts towards the South Galactic Pole versus the  $K_c$ magnitude (green circle) corrected for Galactic latitude (RAVE counts towards both poles are indicated by black crosses,
and RAVE counts of red clump stars are denoted by red plus signs). 
  }
    \label{fig3}
\end{figure} 


\begin{figure}
\begin{center}
\resizebox{8.5cm}{!}{\rotatebox{-90}{\includegraphics{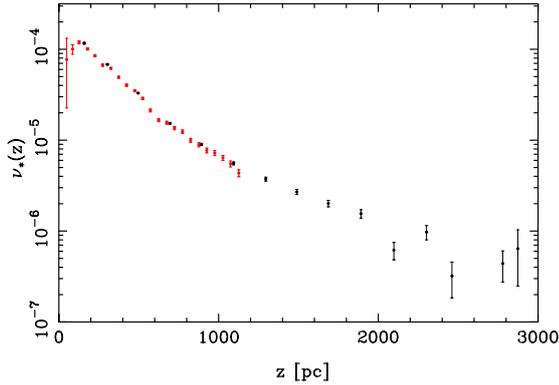}}}
\end{center}
  \caption{Vertical number density distribution of red clump stars towards the South Galactic Pole (black symbols: 200\,pc binning, red  symbols 50\,pc binning.)}
    \label{fig4}
\end{figure}

Figure \ref{fig3} plots the 2MASS counts as a function of $K_c$ magnitudes.     Star counts from our main RAVE sample and from our  RAVE sub-sample of RC stars are also plotted. From these three counts,  we deduce the total number  density of RC stars within the 2MASS sample, as a function of $K_c$, or  equivalently as a function of the height $|z|$ (Figure \ref{fig4}).  Error bars are determined from the three counts   by using the statistical hypergeometric law.

Two  sources of  known bias are present but  remain  small in this analysis.
The first known bias is the degree of homogeneity of the sample selections. Because of  high S/N  ($K<10$),
the accuracy of the various measured or used parameters remains high  independent of  the $z$ distance. For instance, the median accuracy in  $J-K$ colours (within 0.5-0.8)  is  0.03 from $K$=6 to $K$=10. 
Similarly, the mean S/N of the RAVE spectra used to determine the gravity remains high for RC stars at 2\,kpc ($K\sim$10): the mean S/N is 51 (r.m.s 16). This implies that our selections and cuts remain homogeneous independent of the distance $z$.

A second effect is the Malmquist bias: this depends on $\sigma_{\rm M}$, the dispersion of luminosity of the stellar candles, and on the variation of the  density along the line of sight. 
 In the case of a vertical  exponential density law, $\nu \sim \exp(-z/h)$,
  with $h=700$\,pc and $\sigma_{\rm M}$=0.2,
  at  z=1000\,pc
  the bias on the estimated distances is $+2\%$ using a cone for the counts and is $-0.7\%$ using a cylinder.
  At z=2000\,pc
  the bias is +3\% using a cone, and +1.2\% using a cylinder.
 For the   dynamical determination of the  total mass  perpendicular to the Galactic plane, we are  interested in the density gradients, and so just in the variation of this bias: in this study,  it is less than 1\%.
We note that with other tracers with   an absolute magnitude  dispersion  of 0.5,  the bias from star counts would be significantly larger: for cone counts, it is  of the order of 5\% at z=h and 11\% at z=3\,h. This implies a systematic error of  6\% on the resulting determination of the Galactic local surface mass density.

 \subsection{The RC star kinematics}
 
We need to determine the vertical velocities of RC stars that combined with  counts towards the Galactic poles will constrain the vertical potential at the solar position.

Radial velocities are obtained from our RAVE observations, proper motions from the UCAC3 catalogue and distances from the identification of RC stars (see Sections 2.1 and 2.2). 
Radial velocities, proper motions, and distances of RAVE RC stars are converted in (u,v,w) velocities relative to the Sun, and in Galactic velocities,  $V_R-V_{\sun,  R}$ and  $V_z-V_{\sun,  z}$, uncorrected for the solar motion,   assuming $R_0$=8.5\,kpc.
 
\begin{figure}[!htbp]
\begin{center}
\resizebox{8.5cm}{!}{\rotatebox{-90}{\includegraphics{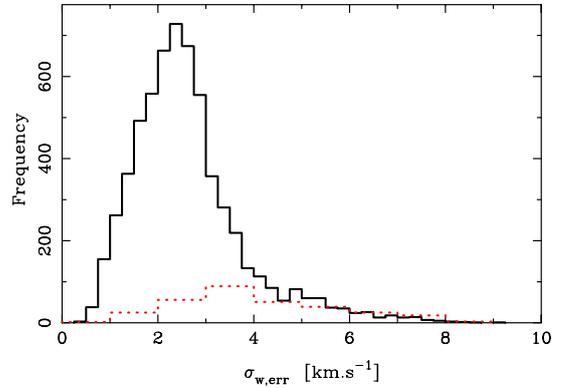}}}
\end{center}
  \caption{ Distribution of errors in the vertical Galactic velocity for stars with $|z|<$2000\,pc (continuous black line), and for stars with $1300<|z|<$2000\,pc (dotted red line).}
    \label{fig5}
\end{figure}

 The errors on the  velocities are obtained from individual  errors on proper motions and radial velocity, adopting a mean uncertainty on distances of 10\%  (Figure \ref{fig5}). The median error on the $V_z$ component is 2.4 km.s$^{-1}$.
 

\begin{figure}[!htbp]
\begin{center}
\resizebox{8.5cm}{!}{\rotatebox{-90}{\includegraphics{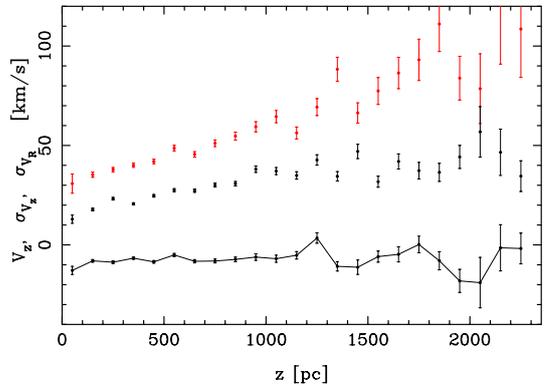}}}
\end{center}
  \caption{ Vertical (black symbols) and radial (red symbols) velocity dispersions: $\sigma_{\rm V_z}$, $\sigma_{\rm V_R}$. Mean  vertical velocity $V_{\rm z}$ (black line).}
    \label{fig6}
\end{figure}

The mean vertical velocity is  constant with $z$ (Figure \ref{fig6}).
The velocity dispersions $\sigma_{\rm R}$ and $\sigma_{\rm z}$ are measured by  applying a 3.5-sigma-clipping to the $V_{\rm R}$,  $V_{\rm z}$  Galactic velocity components. The uncertainties on the dispersions are $\sigma/\sqrt{n_*-1}$.
The vertical velocity dispersion $\sigma_{\rm V_z}$ rises  up to  38 km\,s$^{-1}$ at 1\,kpc and then remains nearly constant (Figure \ref{fig6}).
 
 The velocity ellipsoid tilt is null at  z=300\,pc and reaches 8$\pm$1$\deg$ at 1\,kpc, pointing not far off the Galactic centre. This is in agreement with the finding by \citet{sie08} and \citet{pas12a,pas12b} based on a previous release of the RAVE survey. As discussed in \citet{sie08} a bias on the measure of the tilt exists if no corrections are applied to consider the  anisotropy in the errors of radial velocities and tangential velocities. This bias increases with distance (and with errors on tangential velocities), small at $z$=1kpc, it is about $7\deg$ at $z\sim$2\,kpc.
 Unbiased measurements below 1\,kc indicate that the tilt points not far from the Galactic centre.
 This is confirmed by recent measurements \citep{bud14} based on a large sample of SDSS/SEGUE G dwarfs for which the tilt increases with $z$ and points in a direction close to the Galactic centre. This implies a correlation between radial and vertical motions, which we model using St\"ackel potentials (see Section 3).

\subsection{Metallicities}

To improve the analysis of the vertical potential and the $K_z$ force, we split our sample according to the metallicity in three sub-samples delimited by the values [M/H]=$-$0.35 and $-$0.15. They contain  2182, 2558 and 2263 stars, respectively, of which  1440, 1741 and 1447 are RC stars. The $K_z$ is mainly sensitive to the gradients of the density,  and thus each sub-sample can probe more efficiently a different range of $z$-distances, the lowest metallicity sample probing the potential for the largest distances.

On the other side,  the vertical potential determination is poorly determined at low $z$ because of the lack of data below 300\,pc. For this reason, 
we add a sample of Hipparcos and Elodie RC stars (300 stars) towards the North Galactic Pole. These stars were previously      used to constrain the $K_z$ force \citep{bie06,sou08}. This small sample covers distances from 0 to 800 pc.\\

\subsection{Secondary red clump}

The tracer distance measurement is essential for the accuracy of the vertical potential determination. 

The red clump giants span a large range in gravity depending on their metallicity: $\log g$ = 2.08 for the metal-poor, low-mass end and reaches up to $\log g$ = 3 for the high-mass, metal-rich red clump objects \citep{zha00}.

Identification of clump stars from RAVE observations allow us to achieve a remarkable uncertainty of 7\% in distances if we consider the dispersion of absolute  $K$  magnitude for Hipparcos clump stars with the best parallaxes \citep{gro08}, 10\% otherwise. The existence of a secondary clump at slightly lower magnitude concerns higher mass stars (3 solar masses) for which the He burning does not operate in a degenerate gas \citep{gir99}.  It may degrade the accuracy  of our determination of distances. These stars  are not very numerous compared to ordinary clump stars in the solar neighbourhood, they are  younger and should not be present in older stellar populations with large velocity dispersions or $z$ larger than about 300\,pc. 

This is in agreement with two recent findings. The first is the determination from asteroseismology \citep{ste13} of the mass and of the main or secondary clump status for giants. The second is the comparison of histograms of asteroseismologic mass \citep{mig13} in two COROT fields at different low Galactic latitudes. 

Finally, another source of possible bias is related to  binarity,
however, to modify significantly the apparent magnitude, 
 binary systems consisting of giants are needed.  Such systems are extremely rare, less than 1\% \citep{nat12} since the life time on the red giant phase is extremely short, and  the mass of each companion must not differ by more than 0.5\% to have  a binary consisting of two giants.

\section{Methods and models}

To measure the vertical potential at the solar Galactic position up to a vertical distance of 2\,kpc, we use and adapt the century-old method developed by \citet{kap22} and \citet{oor32}. Their method  can be applied when the stellar oscillations through the Galactic plane remain smaller than $\sim$1\,kpc, so the vertical motions are approximately decoupled from the horizontal ones. Thus, below 1\,kpc from the mid-plane, the problem becomes 1D dynamical model. In this case, the vertical distribution of stellar positions and vertical velocities $f(z,w)$  can be written as the sum of isothermal components depending only on the vertical energy:

$$f(z,w)=\Sigma_i \,\frac{c_i}{\sqrt{2 \pi} \sigma_i} \, \exp\left(-\frac{\Phi(z)-\Phi(0)+\frac{1}{2}w^2}{ \sigma_i^2} \right)$$

with the vertical density of the tracer stars as:

$$\nu(z)=\Sigma_i \,c_i \, \exp\left(-\frac{\Phi(z)-\Phi(0)}{ \sigma_i^2} \right).$$

A general solution is provided in Appendix A. This solution may be written in a different, but equivalent  form found by \citet{gar12}.
Such a solution cannot be applied at higher $z$, where the coupling between radial and vertical motions must be considered in order to achieve an accurate measure of the vertical potential. For that purpose, \citet{kui89}  proposed analytic corrections that they   applied to their sample of K dwarfs  for the determination of the vertical force. \citet{sta89} building exact stationary solutions with St\"ackel potentials found  corrections of the order of 10\% at 1\,kpc in comparison with a 1D model.  

The closeness of the true Galaxy potential to a St\"ackel potential is a very commonly used feature. For instance, \citet{Binney2012} used this similarity to evaluate the three actions associated with any orbit, actions which he then uses as the arguments of his distribution functions. Here, we follow the approach of \citet{sta89} and build a Galactic model with a St\"ackel potential for which the Hamilton-Jacobi equation is fully separable, and a phase-space distribution function depending on the three straightforward integrals of motion in such a separable potential. By construction, the distribution function is stationary and solution of the collisionless Boltzman equation, and is used to model the vertical number density and vertical velocity dispersion of our stellar samples.

To model a realistic  gravitational field within the solar neighbourhood, we follow the mass modelling proposed by \citet{bat94} and \citet{fam03}, using a combination of  two  \cite{kuz62} components, a disc and a halo \citep[a detailed description of their properties  can be found in][]{dej88}. 
We introduce four free parameters: the mass $M$ and  axis ratios $\epsilon$ of both components. These four free parameters allow us to constrain locally the scale-height of the disc, its local surface density, and  the local volume density of the extended component. Thus, the obtained modelled vertical distribution of the total volume density in the solar neighbourhood can represent  any combination of a dark halo and a vertically extended stellar component. The fourth free parameter (the flattening of the halo component) is adjusted  to impose a  flat rotation curve over an extended range of Galactic radius. The $K_z$ fit is thus just made on three parameters. We note that adjusting the flattening of the halo in this way does not modify its local density gradient, the density remaining nearly constant between $z$=0 and 2\,kpc, since the halo is never highly flattened. The halo flatness nevertheless affects the (poorly known) value of the corresponding circular velocity at the Sun's radius.
 
Finally, with St\"ackel potentials, we can very simply model the tilt of the velocity ellipsoid above the Galactic plane. The tilt orientation is fully determined by the  positions $\pm z_0$ of the foci along the vertical axis, $z_0$ defining a confocal ellipsoidal coordinate system. We set $z_0$=2\,kpc, in order to have a velocity ellipsoid at ($R$, $z$)=(8.5\,kpc, 1\,kpc) pointing close to the Galactic centre
 in agreement with observations  \citep{sie08,bud14}.

\subsection{Kuzmin-Kutuzov  potentials}

 A detailed description of all characteristics of 3D St\"ackel potentials  can be found in  
 \citet{dez85}.  These  potentials are easily tractable  in confocal
 spheroidal coordinates.  The prolate spheroidal coordinates ($\lambda,
 \theta, \nu$) are related  to the cylindrical coordinates  ($r,\theta,
 z$) by:

 \begin{equation}
 r^2=\frac{(\lambda+\alpha)(\nu+\alpha)  }{ \alpha-\gamma}~~{\rm  and}~~~
 z^2=\frac{(\lambda+\gamma)(\nu+\gamma) }{ \gamma-\alpha}\, .
 \end{equation}
The shape  of the coordinate surfaces is determined by $\alpha$ and $\gamma$
 while   $\lambda,\,\nu$   satisfy $-\gamma\le\nu\le-\alpha\le\lambda$.
 Surfaces  of  constant $\lambda$  are spheroids  and those of constant
 $\nu$ are  hyperboloids.  They all share  the same foci located on the
 $z$ axis at $\pm z_o=\pm(\gamma-\alpha)^{1/2}$.  
 
 A general St\"ackel  potential takes the form:
 \begin{equation}
 \Phi(\lambda,\nu) = -\frac{ h(\lambda)-h(\nu) }{ \lambda - \nu}
 \end{equation}
 where $h$ is an arbitrary  function.
 
Here, we define  the class of Kuzmin-Kutuzov potentials \citep{dej88}, writing 
 $h(\tau)=GM\sqrt{\tau+q}$ (with the condition $q\ge \gamma$): 
 \begin{equation}
 \Phi_{KK,q}(\lambda,\nu) =-\frac{GM}{\sqrt{\lambda+q}+\sqrt{\nu+q}}.
\end{equation}

The corresponding isodensity surfaces from Poisson equation are flattened oblate spheroids. Increasing $\epsilon^2=(q-\alpha)/(q-\gamma)$ flattens the spheroids. The key is that adding multiple St\"ackel potentials of this type still gives a St\"ackel potential as long as the focal distance $\sqrt{\gamma-\alpha}$ remains the same.

 Besides the energy and the angular momentum, a third independent isolating integral of the motion exists and can be written as:
 \begin{equation}\begin{split}
 I_3 = \Psi(\lambda,\nu)&-\frac{1 }{ 2}\frac{z^2}{ \gamma-\alpha}
 (\dot{r}^2+(r\dot{\theta)}^2) \\
 &-\frac{1 }{ 2}(\frac{r^2 }{ \gamma-\alpha}+1)\dot{z}^2
 + \frac{rz\dot{r}\dot{z} }{ \gamma-\alpha}
 \end{split}\end{equation}
 with
 \begin{equation}
 \Psi(\lambda,\nu)=\frac{ (\nu+\gamma)h(\lambda)-(\lambda+\gamma)h(\nu)
 }{  (\gamma-\alpha)(\lambda-\nu)}.
 \end{equation}

\subsection{The distribution function}

To model the density and kinematics of our samples, we  define a stationary distribution function that depends on  three integrals of the motion. We use the 3D stellar disc distribution function of \citet{bie99}, which has nearly
 a  Schwarzschild distribution   behaviour  in  the  limit   of  small velocity  dispersions.  This  distribution is a generalization of the Shu distribution function \citep{shu69} and also has a density that  is nearly radially exponential.

 The distribution function is
 \begin{equation}
 \begin{split}
 f(E,L_z, I_3) =
 \frac{2\Omega(R_c)}{ 2\pi\kappa(R_c)}
 \frac{\Sigma(L_z)}{
            \sigma_r^2(L_z) }
             \exp \left[
 -\frac{E-E_{circ}}{\sigma_r^2}
 \right] \\
 \frac{ 1}{ \sqrt{2\pi} } \frac{1}{\sigma_z(L_z) }
 \exp \left\{
 -\left(\frac{R_c(L_z)^2 }{ z_o^2 } +1 \right)^{-1}
 \left(\frac{1}{\sigma_z^2}-\frac{1}{\sigma_r^2}\right)I_3
 \right\} 
 \end{split}
 \end{equation}
  
 with $R_c(L_z)$  the radius of the circular orbit that has the angular momentum $L_z$, 
 $\Omega$ is the angular velocity, $\kappa$ is the epicyclic frequency, and $E_{circ}$
is the energy of a circular orbiting star at radius $R_c$.  

For sufficiently small velocity dispersions, the number density distribution,
$  \Sigma(L_z) =\Sigma_0 \exp( -R_c/R_{\nu})$,
 is close to $  \Sigma(R) =\Sigma_0 \exp( -R/R_{\nu})$.  

We set $  \sigma_{r,z}(L_z) =\sigma_{0,r,z} \exp( -R_c/R_{\sigma_{r,z}})$
and the velocity dispersions are close to
$  \sigma_{r,z}(R) =\sigma_{0;r,z} \exp( -R/R_{\sigma})$. 

Local number density and dispersion, $\Sigma_0$, $\sigma_0$, are constants,  $R_\nu$ is (close to) the scale length of the number density distribution, and $R_\sigma$ is (close to) the scale length of the velocity dispersions.

This distribution function is very  similar to that proposed by \citet{sta89}, but here, generalized to the cases where $R\ne R_0$. We also  reduced the number of free parameters to have a   form  closer to the \citet{shu69} distribution function.
The corresponding velocity distribution is  not far from a 3D gaussian. When $z$=0, the  corresponding density varies nearly exponentially in an extended domain of a few kpc around the Sun.  We note that if $z$$\ne$0, the density above the plane also depends on the vertical potential; thus the density may vary exponentially  at any $z$  with a supplementary condition, as for instance $R_{\sigma}= 2R_\rho$   \citep[see also eqs.~6-7 of][]{kru01}. The velocity dispersions also decrease exponentially.
This distribution function had been previously used for a dynamically consistent analysis of the kinematics of Hipparcos stars  \citep{bie99}.

For the modelling of the distribution function of our RC star samples, we fix the scale lengths for the radial density and for the velocity dispersions: $R_\nu$=2.5\,kpc and $R_{\sigma}$=9\,kpc.
 This is to be compared with 
$R_\nu$\,$ \sim$\,2.2 and 2.8\,kpc for the thin and thick discs 
\citep{cab05,jur08, cha11, pol13, rob14}
 and $R_{\sigma^2}$=4.4 or 5.6\,kpc, ($R_{\sigma}$=8.8 or 11.1\,kpc), respectively by \citet{lew89} and \citet{ojh96}.
We also  set 
$\sigma_r / \sigma_z $ =2
for the thin disc stars ($\sigma_z < 27$km.s$^{-1}$)  and 1.5 otherwise, in agreement with the observed properties of our RC sample.

\subsection{The corrected bias}

The introduction of a 3D model allows us to correct different effects relative to a 1D vertical model.
The first effect is the velocity ellipsoid tilt at large $z$ that increases the observed vertical velocity dispersion.
The second effect is related to the vertical bending of the stellar orbits. For stars seen at $z$=1 or 2\,kpc, their mean Galactic radius $<R>$, when they cross the Galactic plane, is  larger than $R_0$, a position where the stellar density is lower than at the radius $R_0$. These two effects lead to an overestimate of $K_z$ in a 1D model. A third effect is  also related to the bending of orbits and to the radial gradient of $\sigma_z$, lowering $\sigma_z$ towards the pole. This effect leads to an underestimate of the $K_z$ force in a 1D model.

Here, the modelling with a locally valid St\"ackel potential allows  us to correct in a dynamically consistent way the  bias resulting from the coupling of vertical and horizontal stellar motions. However, this is  obtained  at the expense of supplementary parameters: the tilt orientation of the velocity ellipsoid, the radial gradients of the stellar density, and of the kinematics.

\subsection{The adjustment procedure.}

We determine the parameters of the vertical potential by fitting the observed moments, density $\nu(z)$ and vertical velocity dispersions $\sigma_{zz}(z)$ for each of the three metallicity samples; moments are computed from the distribution function Eq. 8. We minimize the difference between observed and modelled quantities using the $\chi^2$:

$$
\chi^2=\chi_\nu^2 + \chi_\sigma^2
+\mathrm{q} \left( \,V_c(9.5kpc) -V_c(7.5kpc) \, \right)^2
$$
with
$$
\chi_\nu^2=\Sigma_i \frac{\left( \nu_{mod,i}-\nu_{obs,i} \right)^2}{\epsilon_{\nu,i}^2}
$$
$$
\chi_\sigma^2=
\Sigma_i \frac{\left( \sigma_{mod,i}-\sigma_{obs,i} \right)^2}{\epsilon_{\sigma,i}^2}\,,
$$

where the $\epsilon_i$  are the corresponding uncertainties.

For the density, the bin size is 100\,pc between 300\,pc and $\sim$2000\,pc.
We do not consider our  density estimates below 300\,pc where the completeness is difficult to determine. For the velocity dispersions, the bin size is 100\,pc between 200\,pc and 1200\,pc and 200\,pc beyond.

The last r.h.s. term within the $\chi^2$ expression  is introduced to impose a potential with a nearly flat  rotation curve.
The $q$ factor does not need to be large and the contribution of the corresponding term to the $\chi^2$ is small, because the slope of the rotation curve  is uncorrelated to the other parameters.  The adjusted rotation curve is flat and varies by  1\,km\,s$^{-1}$ between $R$=7 and 15\,kpc.

We use two quasi-isothermal components (Eq. 8) to model the stellar sample distribution function, the density $\nu(z)$ and dispersion $\sigma_{z}(z)$ for each metallicity sample.\\

We  determine the $\chi^2$  minimum for the parameters of the potential using  the  MINUIT software \citep{jam04} that allows us to look for possible multiple minima and  to  obtain a first estimate of the variance-covariance matrix. To compute the posterior probability distribution function (PDF), we  consider the  likelihood ${\cal L}=\exp(-\chi^2/2)$ and  apply a Markov chain Monte-Carlo using the Metropolis-Hastings Algorithm \citep{for13}. From this, we determine the marginal PDFs.


\begin{figure}[!htbp]
\begin{center}
\resizebox{8.5cm}{!}{\rotatebox{-90}{\includegraphics{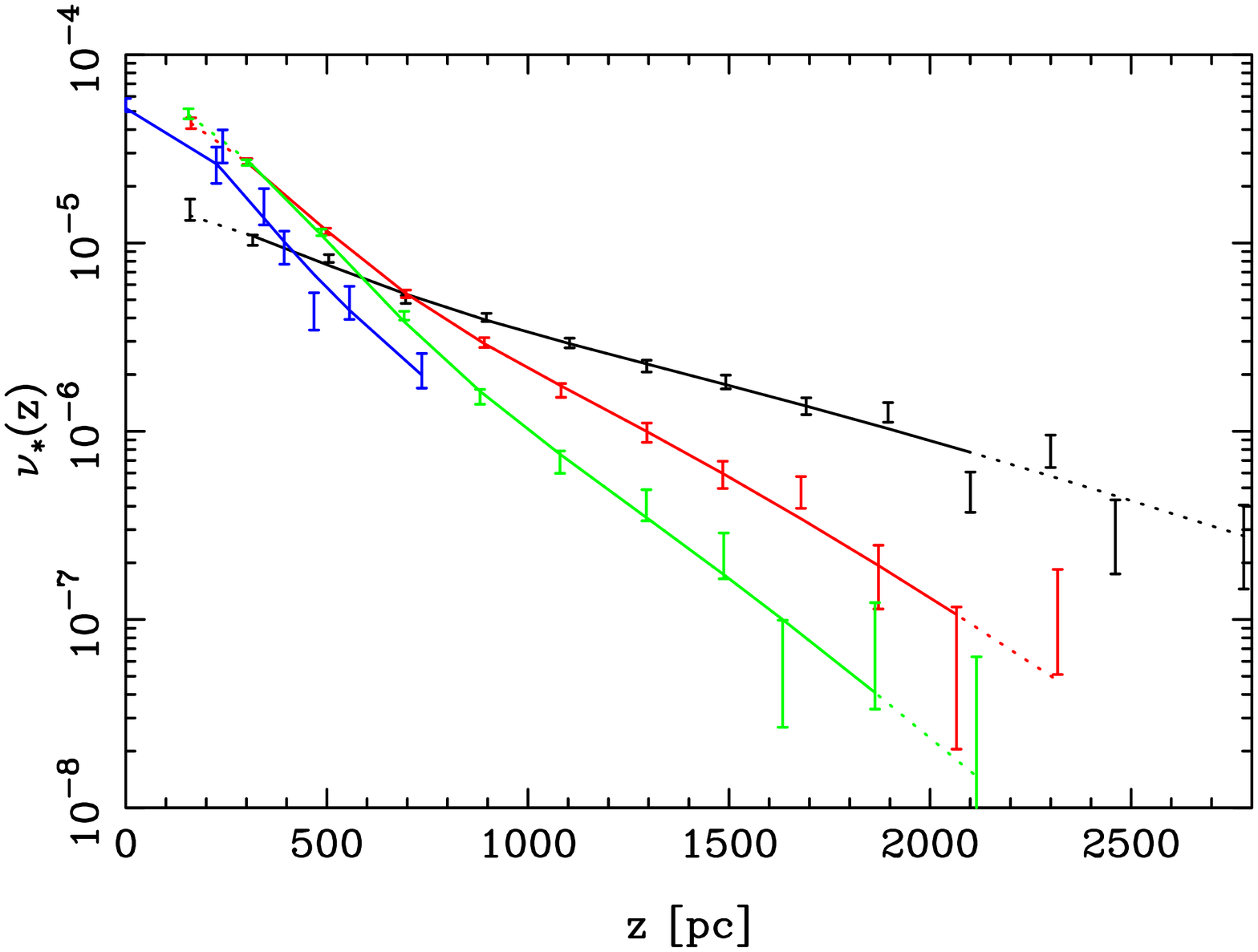}}}
\resizebox{8.5cm}{!}{\rotatebox{-90}{\includegraphics{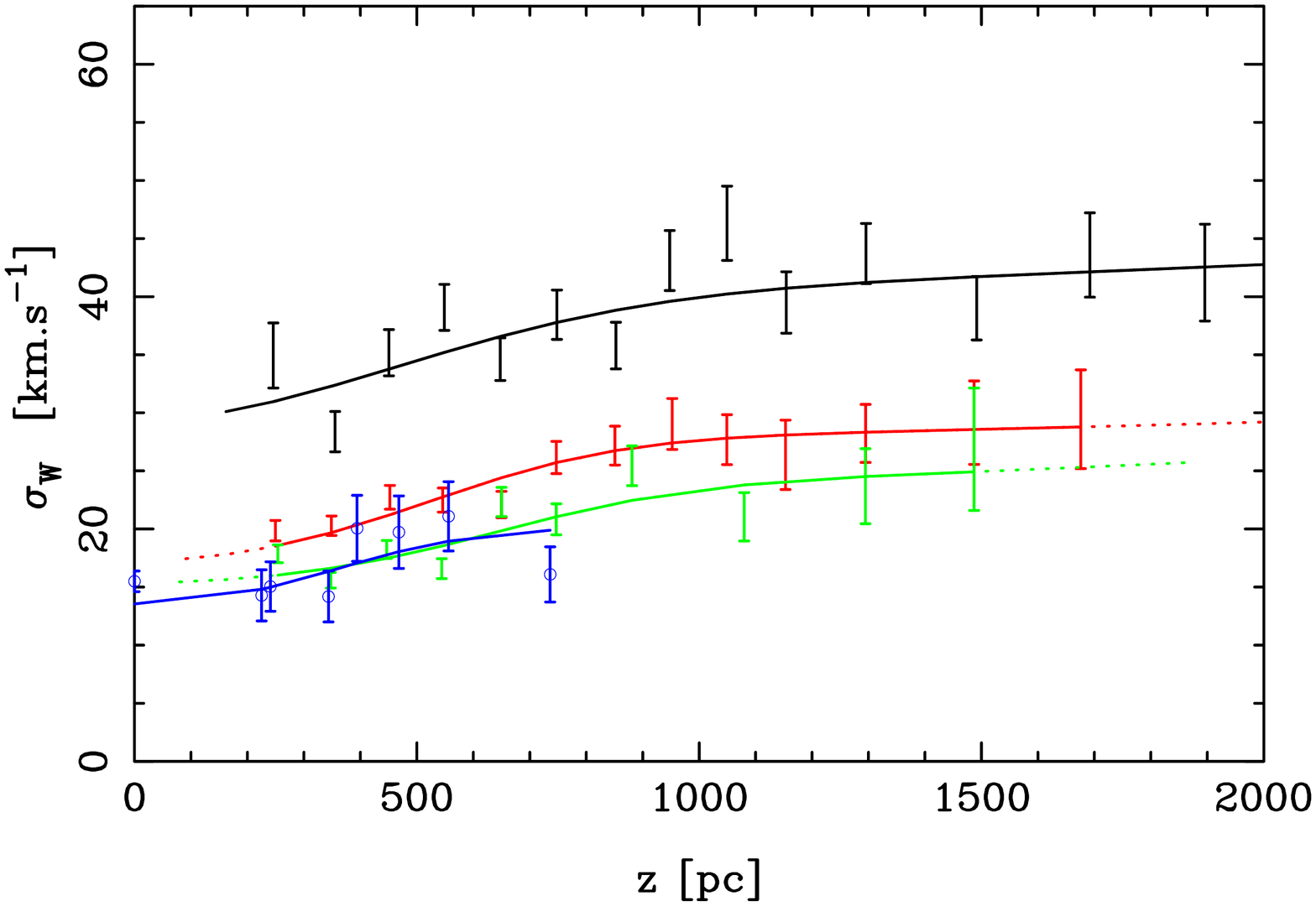}}}
\end{center}
  \caption{Vertical number density (top) and velocity dispersions (bottom), data (symbols with error bars) and model (lines) for the three RAVE metallicity samples and the NGP-Hipparcos sample: low metallicity $[M/H]\le-0.35$ (black), intermediate  $-0.35<[M/H]\le-0.15$ (red), high  $-0.15<[M/H]$(green), NGP (blue). Dotted lines are the extrapolated model not fitted to data.
  }
    \label{fig7}
\end{figure}


\begin{figure}[!htbp]
\begin{center}
\resizebox{8.5cm}{!}{\rotatebox{-90}{\includegraphics{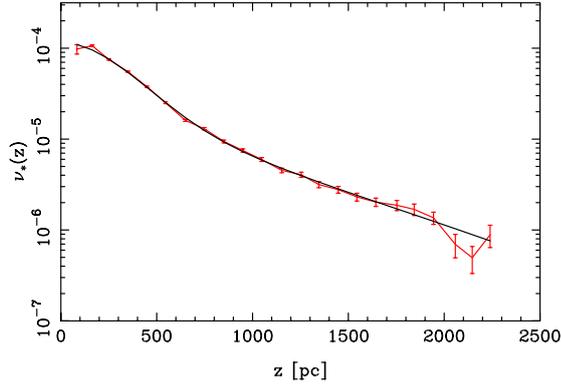}}}
\resizebox{8.5cm}{!}{\rotatebox{-90}{\includegraphics{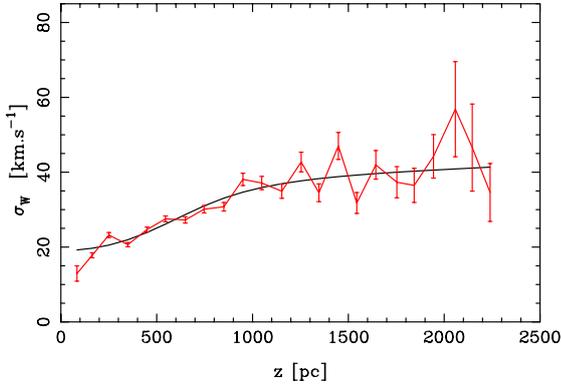}}}\end{center}
  \caption{Vertical number density (top) and velocity dispersions (bottom) for the full RAVE RC sample (symbols with error bars) and model (continuous line).
  }
    \label{fig8}
\end{figure}

Figure \ref{fig7} shows the binned data and the best fit model to the density and dispersion versus $z$ for the three metallicity RAVE samples and for the Hipparcos-NGP sample. For information, we also show this  best-fit model  and the binned data for density and vertical velocity dispersions  without metallicity splitting (Figure \ref{fig8}).

\section{Results}

\subsection{The $K_z$ force}

In the previous sections, we have described the stellar samples used to probe the vertical force towards the Galactic pole. Our method does not fundamentally differ from the century-old pioneering work of \citet{oor32}. The modelled quantity that is fitted to the observations is the vertical potential within the interval of distances probed by our stellar tracers. The vertical force and the total  mass density distributions are deduced  from the first and second $z$-derivatives  of this potential.  It is known that the $K_z$  is an ill-conditioned problem and that without a filtering of data or smoothing assumptions about the shape of the potential, the derivatives would be dominated by the noise and fluctuations of data. 
Here, the actual smoothing assumption for the local potential is given by two Kuzmin-Kutuzov components \citep{bat94,fam03} to mimic the potential of a disc and a spheroid (and also a flat rotation curve). This is partly equivalent to the three-parameter modelling introduced by \citet{kui89} to describe the total Galactic  disc mass surface density. Our model   goes beyond the plane-parallel approximation and allows us to describe the distribution function $f(z,w)$  beyond 1\,kpc up to 2\,kpc. The vertical potential is modelled by a disc, its local density and thickness, and by the local density of an extended component that represents the dark matter halo.


\begin{figure}[!htbp]
\begin{center}
\resizebox{8.5cm}{!}{\rotatebox{-90}{\includegraphics{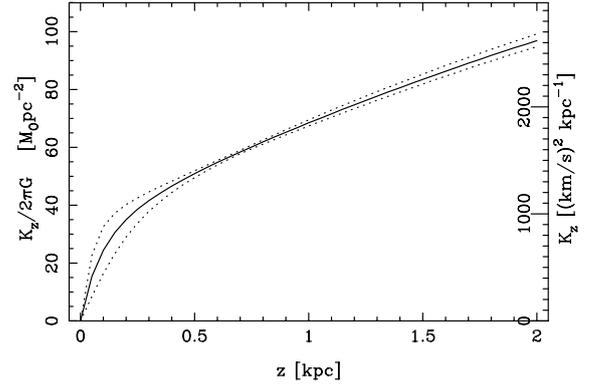}}}
\end{center}
  \caption{  The $K_z$ force and 1-$\sigma$ error intervals. }
    \label{fig9}
\end{figure}

The resulting $K_z$ at $R_0$ from the fit of  RAVE data (Figures \ref{fig7}) is shown in Figure \ref{fig9}. We give the results in terms of $K_z/2 \pi$G for visibility and comparison with other studies, even though this should not be confused with the true surface density.

The obtained $K_z$ is less constrained at low $z$ where there are no RAVE data, but only the  smaller sample of Hipparcos and Elodie stars.  The $K_z$ force  at 350\,pc is  found to be 44.2$^{+2.3}_{-2.9}\,$M$_{\sun} $pc$^{-2}$ in agreement  with the \citet{kor03} determination $42\pm6\,$M$_{\sun} $pc$^{-2}$, based on a sample of 1500 K giants from Hipparcos observations, mainly first ascent giants rather than clump giants.

We also note that our $K_z$ determination near $z$=0 allows us to deduce the Oort limit $\rho_{\rm dyn}$(z=0)=0.0911$\pm0.0059\,$M$_{\sun} $pc$^{-3}$, intermediate value between   the \citet{cre98} (0.076$\pm0.015\,$M$_{\sun} $pc$^{-3}$) and \citet{hol04} (0.102$\pm0.010\,$M$_{\sun} $pc$^{-3}$) determinations. These two studies are based on the same Hipparcos data, below $z$=125\,pc, but with different assumptions regarding the shape of the potential.  Most likely, this  explains  the difference between both results, which  differ by  just a 1-$\sigma$ error.

Our  $K_z$ force determination from 0 to 1\,kpc, is  similar  to recent studies, but in our case, with more free parameters and without limiting assumptions on the baryonic local surface density or  on the dark matter local volume density. 

Since the quasi-totality of the ordinary matter  resides below $z$=1\,kpc, the mass density beyond  1\,kpc is dominated by the dark matter. This results that  our  $K_z$ measurement gives  direct access to the dark matter density between 1 and 2\,kpc above the Galactic plane. 

The $K_z$ force at intermediate and high $z$ distances are:

   \noindent $K_z$(1kpc)/(2$\pi$G) =68.5$\pm$1.0$M_{\sun}$pc$^{-2}$ =1852$\pm$27 km$^{2}$ s$^{-2}$ kpc$^{-1}$, and 
   
  \noindent   $K_z$(2kpc)/(2$\pi$G) =96.9$\pm$2.2$M_{\sun}$pc$^{-2}$ =2619$\pm $59 km$^{2}$ s$^{-2}$ kpc$^{-1}$ .
 
\subsection{ Dependence on fixed model parameters} 

The vertical tilt of the velocity ellipsoid is fixed in our models to point close to the Galactic centre through the choice of our foci 
\citep[tilt of $\sim$13\,$\deg$ at $z$=2\,kpc, in accordance with][]{bud14}.
 Unbiased measurements indicate that the tilt indeed points close to the Galactic centre, and our method is not affected by the existence of a bias in the tilt that one would actually measure with RAVE data at large heights (Sect. 2.3) because we directly fit individual velocities (Sect. 3.4) and not the global shape of the velocity ellipsoid.

The two other fixed parameters, non-existent in traditional analysis assuming separation of vertical and radial motions, are  the  radial scale lengths of the number density of tracer stars and of the velocity dispersions.
 
The radial scale length of the stellar sample, $R_\nu$,   modifies by less than 1\% the $K_z$ between $z$= 0 to 2 kpc, when $R_\nu$ is varying from 1.8 kpc to 3.5 kpc. Increasing the kinematics scale length does not modify  the $K_z$ determination. Only  decreasing the scale length $R_\sigma$ to 7 or 5 kpc (or $R_{\sigma^2}$  to 3.5 or 2.5\,kpc)   increases the $K_z$ force at 2 kpc by 5\%
and 11\%, respectively. This implies that our determination of  the surface mass density of disc would be  be increased by 5\% or 10\%, and the local DM density  by 5\% or 14\%. However, these small kinematic scale lengths are excluded by existing observations.

\subsection{The vertical mass density}

\begin{figure}[!htbp]
\begin{center}
\resizebox{8.5cm}{!}{\rotatebox{-90}{\includegraphics{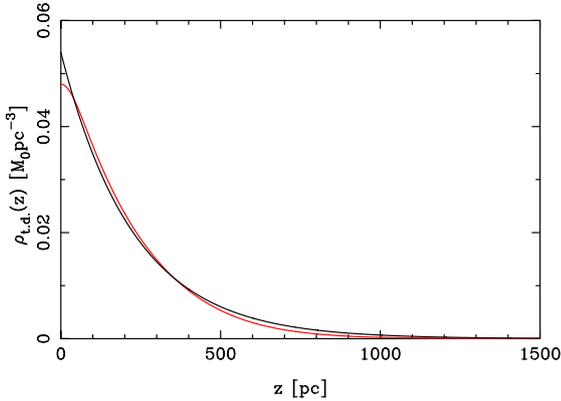}}}
\end{center}
  \caption{  Vertical mass density of the stellar thin discs from the Besan\c con model (red line) fitted with an exponential of scale height $h$=229\,pc (black line).}
    \label{fig10}
\end{figure}

To be able to estimate the vertical mass density distribution of Galactic components, the $K_z$ determination is not completely sufficient, and  we must also know the 3D shape of the baryonic and dark matter components. Here, the four-parameter St\"ackel potential we fitted should be considered simply as a way to estimate the $K_z$ force itself, but the relative contribution of the baryonic disc and halo to this force can be more reliably dealt with a posteriori by representing the baryonic mass component with a  double exponential  law $\rho(R,z) \sim \exp(-R/H_{\rho}) \exp(-|z|/h_z)$, whose mass is assumed to be proportional to the stellar discs, the dominant baryonic mass component. Recent analyses of Galactic star counts with accurate and detailed modelling of the luminosity functions  converge towards a short scale length, 2.1 to 2.3\,kpc for the thin disc \citep[see for instance][]{rob14} and between 2.8 to 3.2 for the thick disc, while the vertical density of the stellar disc population is very close to an exponential (Figure \ref{fig10}). 


For a given mass surface density of the disc at $R_0$, both parameters $H_{\rho}$ and  $h_z$ can modify the vertical force. Decreasing the scale length $H_\rho$ increases the vertical force of the disc. In this case, to fit the observed $K_z$,  the surface mass density of the disc must be decreased (and  to  fit the $K_z$ at larger $z$, the dark matter mass density $\rho_{\rm DM}$ must  be increased).

Here, we  will consider as reference values $H_{\rho}$=2200\,pc,  $h_z$=300\,pc and $R_0$=8500\,pc for the double exponential disc.

We assume that  the dark halo component is  spherical. Its radial mass density is defined to exactly complement  the double-exponential disc component in order that our model of the Galactic rotation curve  is strictly flat.

\begin{figure}[!htbp]
\begin{center}\
\resizebox{8.5cm}{!}{\rotatebox{0}{\includegraphics{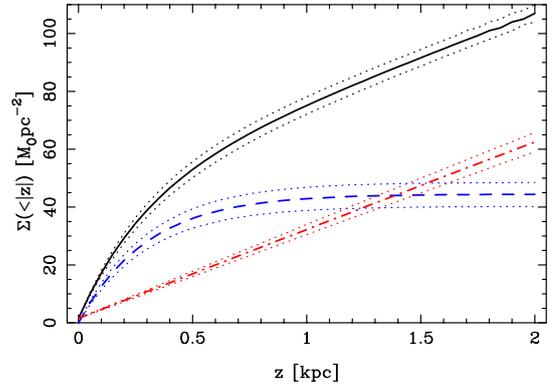}}}
\end{center}
  \caption{  Total surface mass density (black continous line) $\Sigma(<|z|)$=$\int_{-z}^z \rho_{tot} dz$ at $R_0$ split in  a  DM spherical component (red dashed line) and in a baryonic double exponential disc (blue dash-dotted line). Dotted lines are 1-$\sigma$ error intervals. }    \label{fig11}
\end{figure}

Two free quantities remain: the mass or the local density of each component. We adjust  the local density of the baryonic and of the dark matter to fit in a least square sense the observed $K_z$.
The resulting total vertical mass density  distribution is shown in Figure \ref{fig11}  with the DM halo and disc decomposition. The plotted surface densities are the integrated mass volume density between $-z$ and $z$.
We obtain for  the total surface density of the disc component at the solar position:

$\Sigma_{\rm disc}(R_0)=44.4\pm4.1\, $M$_{\sun} $pc$^{-2}$.

\begin{figure}[!htbp]
\begin{center}\
\resizebox{8.5cm}{!}{\rotatebox{-90}{\includegraphics{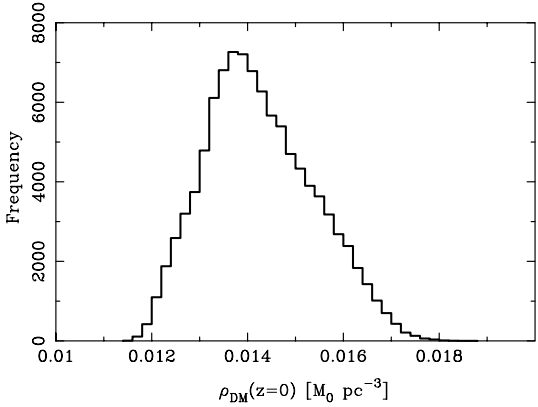}}}
\end{center}
  \caption{  Probability distribution function for the local dark matter density $\rho_{\rm DM}(z=0)$.
  }
    \label{fig12}
\end{figure}

\noindent The probability distribution function of total local volume mass density of the dark matter component is plotted in Figure \ref{fig12}. This yields

\noindent $\rho_{\rm DM}(z$=0)=0.0143$\pm$0.0011\,$ $M$_{\sun} $pc$^{-3}$
=\,0.54$\pm$0.004\,${\rm Gev\, cm^{-3}}$.

In the case of the baryonic and the total (baryonic+DM) local volume densities, the Oort limit, we obtain 

\begin{figure}[!htbp]
\begin{center}\
\resizebox{8.5cm}{!}{\rotatebox{-90}{\includegraphics{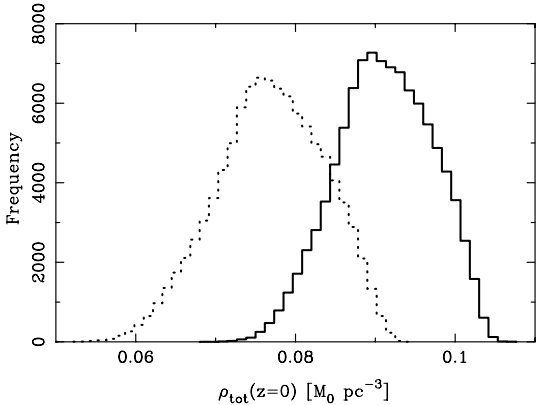}}}
\end{center}
  \caption{  Probability distribution functions for the  total local mass density, i.e. the Oort limit $\rho_{\rm tot}(z=0)$ (continous line), 
  and the baryonic  mass density $\rho_{\rm bar}(z=0)$ (dotted line).
  }
    \label{fig13}
\end{figure}

$\rho_{\rm baryons}(z$=$0)=0.077\pm0.007\, $M$_{\sun} $pc$^{-3}$,

$\rho_{\rm total}(z$=$0)=0.091\pm0.0056\, $M$_{\sun} $pc$^{-3}$,

and Figure \ref{fig13} plots their respective probability distribution function.


Table\,1 shows the results for  $\rho_{\rm DM}$ and $\Sigma_D$ in the case of some other  disc parameter values (results are based on a  $\chi^2$ minimization and are slightly different from these resulting from   MCMCs).
The change of the best-fit solution with the chosen method could result from the imperfect adequacy of the model. This would indicate a bias; this dependency of the results on the adopted methods  has been also shown in a different context by \citet{pol13} analysing  Galactic star counts.

\begin{table}[htdp]
\caption{Disk-halo parameters  reproducing the observed $K_z$ vertical force. The resulting level of the flat rotation curve, $V_c$, is added for information. }
\label{table1}
\centering
\begin{tabular}{ c c c c c c c c c c }
\hline\hline 
$R_0$ & $H_\rho$ & $h_z$&  $\Sigma_D$ & $\rho_{d.m.}$ &  $V_c$\\
\hline
pc & pc& pc& $M_{\sun} pc^{-2}$ & $M_{\sun} pc^{-3}$ & km.s$^{-1}$\\
\hline
8500 & 2200& 300 & 45.6 & 0.0154  & 267\\

8500 & 2000 & 300 & 45.0 & 0.0157 & 263\\

8500 & 2400 & 300 & 46.5 & 0.0146 & 264\\

8500 & 2200& 250 & 43.2 & 0.0158 & 268\\

7500 & 2200 & 300 & 66.4 & 0.0166 & 263\\
\hline
\end{tabular} 
\end{table}

We can note that, remarkably, the dark matter density is nearly insensitive to the model parameters $H_\rho$, $h_z$, $R_0$. This results from the fact that the dark matter model has a nearly constant density in the range z=0 to 2 kpc and  depends quite exclusively on the difference on the $K_z$ force between 1 and 2 kpc. 

 \section{Discussion}

The last RAVE data release \citep[DR4, ][]{kor13} allows us to probe the vertical density distribution of RC stars to a distance of 2\,kpc from the Galactic plane, and  also to determine  their vertical kinematics and metallicity. This provides a highly accurate sample for the study of the vertical force perpendicular to the Galactic plane. About 5000 RC stars are  used, permitting us to relax some of  the traditional assumptions. Specifically, because of the large range of vertical distances  probed up to 2\,kpc,  we can separate the contribution to the vertical force due to the halo and   the disc.

The $K_z$ problem is, in principle,  ill-conditioned. The potential is the fitted quantity, and to determine the total mass vertical density,  we  evaluate the second derivative of  the potential. To avoid arbitrary fluctuations resulting from the  finite size of our stellar tracers, it is necessary to smooth the potential  to  recover a realistic vertical density. Here, this is done by assuming the local vertical potential shape as a St\"ackel combination of a disc and a spheroidal halo. This is accomplished with a three-parameter model constraining the local densities of a disc and of a spheroidal halo, and the thickness of the disc. A fourth parameter imposes a flat rotation curve. The flatness of the rotation curve does not directly impact the $K_z$ determination, but allows us a more realistic  dynamical self-consistent distribution function modelling the density and velocity distribution of tracer stars.

\subsection{The $K_z$ force measurement}

At $z$=1\,kpc, our $K_z$ measurement is similar and in  agreement with the two last decades determinations of the $K_z$ force based on data from stars up to  $z$$\sim$1\,kpc:
$K_z(1\,$kpc/2$\pi$ G)$\sim$68\, M$_{\sun}$ pc$^{-2}$
\citep[for recent determinations, see][]{gar12,bov13,zha13}. 
Between $z$=1 and 2\,kpc, we find that the scale-height of the dark halo is significantly larger than the domain probed with our sample, and that the density is nearly constant below $z$=2\,kpc.

The local density of the halo is $\rho_{\rm DM}(z$=0)= 0.0143$\pm$0.0011M$_{\sun}$\,pc$^{-3}$=0.542$\pm$0.042\,{\rm Gev\,cm$^{-3}$}. 
We also determine the total surface density of the disc components 
$\Sigma_{\rm bar}$=44.4$\pm$4.1\,M$_{\sun}$ pc$^{-2}$
that is in agreement with the \citet{fly06} determination.
The halo volume density and disc surface density are determined nearly independently, with the consequence that their resulting uncertainties are small. Nevertheless, systematic uncertainties, e.g. due to the choice of smoothing of the potential, are, in fact, larger.

 Most of the recent determinations claimed estimates of the local halo density  to be of the order of 
 $\rho_{\rm DM}$(z=0)=0.006-0.008\,M$_{\sun}$ pc$^{-3}$
  \citep[see the compilation in Table 4 by ][]{rea14}
However, it must be noted that
in these previous  $K_z$ measurements, the $z$ extension of the stellar tracers was limited to distances of 0.8 to 1.1\,kpc. For that reason, it was not possible to accurately separate the respective contributions from the visible discs (stellar and ISM discs) and those contributions from the dark matter halo. Then  independent estimates  of the visible matter from star counts and kinematics were adopted.
As  mentioned in 
the review by \citet{rea14}, these determinations of the local dark matter density were based on  an assumed total surface disc density for the visible matter. 
In one study, \citet{kui89} assumed the value of the density of the dark matter  and deduced the surface mass density of the disc.

\subsection{The local volume mass density determination}

To move from the vertical force to the vertical density distribution and 
to the decomposition  in  contributions from   disc and from halo components implies we know  non-local Galactic characteristics
as the scale length of the  disc, the Galactic solar radius $R_0$, or  the thickness of the visible matter.
Unfortunately, many fundamental characteristics of our Galaxy's structure remain inaccurately measured, such as the distance of the Sun to the Galactic centre, or  are even subject to contradictory determinations, such as the amplitude and shape of the Galactic rotation curve. For this reason, we  list  the consequences of our findings of high local density of dark matter, according to different hypotheses and to results recently published concerning the Galactic gravitational potential.

First, the
large scale and 3D properties of the dark matter distribution are mainly established from  the knowledge of the Galactic rotation velocity curve or  from orbits of streams through the Galactic halo. Other constraints exist locally, for instance the Galactic escape velocity in the solar neighbourhood \citep{smi07,pif14} or the determination of  Oort's constants \citep{oll98}.

The flatness of the Galactic rotation velocity curve at $R_0$ is consistent with  the observations of external disc galaxies of  similar type. The flatness  is  supported by the two recent determinations of the Galactic circular velocity curve from parallaxes \citep{rei14} or from spectro-photometric distances \citep{bov12} that  favour a flat rotation curve in an extended interval  of radii around $R_0$.

Now, considering such a  flat rotation curve with $V_c$$\sim$220\,km\,s$^{-1}$, similar to the recent determination from the APOGEE project \citep{bov12}, and a spherical dark matter halo  (frequently called the ``standard dark Galactic halo model'' within the astroparticles literature), \citet{sal10} and others  built Galactic models and estimated $\rho_{\rm DM}$ $\sim$0.006-0.01 M$_{\sun}$\,pc$^{-3}$ in the solar neighbourhood \citep[see Table 4 from][]{rea14}. 

If we consider the Galactic rotation curve \citep{rei14} deduced from data of the BeSSeL project \citep{bru11}, which 
 leads to a higher circular velocity curve of about $V_c$$\sim$240 km\,s$^{-1}$\citep{mcm11,bob13,rei14}, 
 and also  assuming a spherical dark matter halo, \citet{mcm11} obtained $\rho_{\rm DM}$=0.40$\pm$0.04\,Gev cm$^{-3}$(=0.010\,M$_{\sun}$/pc$^{-3}$) a  relatively low value still.

Concerning the local slope of the rotation curve, some of  the strongest evidence comes from the determination of Oort's constants  by \citet{oll03}, based on Hipparcos data \citep[see also][]{mig00}. Their determination is based on the oldest  red giant stars, over a large domain of 2\,kpc radius around the Sun. Their careful analysis considers  the necessary corrections for the bias due to the extinction and also to the mode mixing from the solar motion. They found out that the rotation curve is flat over $\pm$\,2\,kpc from the Sun's Galactic radius, varying by less than $(A+B)/2= dV_c/dR$=$+0.5\pm0.8 $ km s$^{-1}$ kpc$^{-1}$ (and $d\ln V_c/d\ln R$=+0.02). Their value of the slope of velocity curve introduces a small additional  non-zero contribution to the local mass density  from the Poisson equation:
$$(4 \pi G R )^{-1}\frac{\partial V_c^2}{\partial R}=
\frac{B^2-A^2}{2\pi G}
=+0.0012\pm0.0019~{\rm M}_{\sun} {\rm pc}^{-3}\,.$$

For high values of the local dark matter density such as those we obtained here, we can mention the work by \citet{bur13} who  built a global dynamical Galactic model to constrain $\rho_{\rm DM}$ at the Galactic centre and at $R_0$  from published observations of the circular velocity curve. 
They  plot the $K_z$ force that  can  be directly compared with our measurement. One of their $K_z$ models (figure 6b) is in  agreement with our determination
within $z$=0 to 2\,kpc and they obtained 
$\rho_{DM}(z=0)=0.015\,$M$_{\sun}\,$pc$^{-3}$. Unfortunately, in their model they adopt a rapidly rising rotation curve at $R_0$ that we  judge very unlikely. This rising rotation curve, with $d\ln V_c/d\ln R\sim$0.16 explains their high value obtained for $\rho_{\rm DM}$. In their model, if the rotation curve was flat,  the local density would   probably be about 0.006\,M$_{\sun}\,$pc$^{-3}$. 

In a recent study, using tracer stars with $z$ between 1 to 2\,kpc,  \citet{smi12} estimated an order of magnitude for  $\rho_{DM}(z=0)=0.015\,$M$_{\sun}$\,pc$^{-3}$,   similar to our finding. However, because of their  crude modelling of the potential and  the lack of a clear definition of the selection function, they decided not to give error bars for their estimate. It remains  remarkable that this unique previous analysis of distant tracer stars agrees with our finding. Finally, \citet{pif14b}  also recently determined a similar value as ours with the RAVE data, for a halo flattening of $0.8$.

\subsection{Global Galactic properties }

When we consider our model   with $\rho_{\rm DM}$=0.014\,M$_{\sun}$\,pc$^{-3}$, a flat rotation curve and a spherical halo, this implies  $V_c$=267\,km\,s$^{-1}$, which is a too large value compared to the recent determinations of the Galactic rotation curve. The simplest way to reconcile our  local determination of the dark matter density   with the admitted flat rotation curves based on direct observations  consists in flattening the dark matter halo.
An axis ratio of the order of 0.67 is necessary, in the case of $V_c$=220\,km\,s$^{-1}$ as determined by \citet{bov12}. This significant flattening is not in agreement with the results of Galactic cosmological numerical simulations \citep{mac07} for which a mean flattening of the order of  0.8 is expected. 
 Moreover, \citet{pil14} analysed simulations including dissipational gas physics and obtained much rounder halo with $q$=0.99, instead of $q$=0.53 in the case of DM-only numerical simulations.

In fact, combining the circular velocity from the BeSSeL project $V_c$\,$\sim$\,240\,km\,s$^{-1}$\citep{mcm11,bob13,rei14}, with a nearly flat rotation curve at the Sun position, and a flattening of $0.8$, leads to our estimated value of the local dark matter density, in accordance with  \citet{pif14b}. This value of the circular velocity makes the Milky Way a clear outlier from the Tully-Fisher relation \citep{hol06,ham12}.
 
Another plausible explanation of a high local DM density  comes  from the cosmological numerical simulations by \citet[][figure 9]{rea09}  and \citet{pil14} 
\citep[see also][ for DM detection implications]{lin10}. 
They showed that in the case of a disc galaxy already in place at high redshift, the later accretion of   galaxy satellites create a slowly rotating very thick disc or flattened spheroidal component of dark matter. This dark component  results from the accretion of the dark component of each accreted satellite. Due to the history of accretion, this accreted DM component has a high angular momentum, with kinematical properties intermediate between the stellar disc and a non-rotating spherical dark halo.
Its detailed structure depends  on the details of the accretion history, and the halo mass depends on the unknown number and mass of accreted satellites.
The local contribution of this accreted DM component could be 25 to 150 percent the density of the primordial and  nearly spherical DM component. In the case of a small  scale height  of the order of 2-3\,kpc, its vertical density is quickly decreasing. In such a case, our modelling of the $K_z$ force probably does not include a sufficient number of free parameters to accurately model  the shape of the $K_z$ force between 1 and 2 kpc. We might suspect that our modelling forces a nearly linear rise of the $K_z$ between 1 and 2\,kpc. 
A supplementary (thick disc) Kuzmin-Kutuzov  component could be added \citep{fam03} to model more precisely the vertical dark matter density and potential, but it is not clear that the size of our sample will allow us to discriminate between models with so many parameters.

A similar disc of ``phantom'' dark matter (from the point of view of a Newtonian observer) is predicted  \citep{bie09} by the MOND effective theory. This can be a very similar effect to that  observed in numerical simulations   of accretion of satellite galaxies. For a baryonic model like that assumed here, a vertical force $K_z$/(2$\pi$G)$\,\sim$ 90\,M$_{\sun}$\,pc$^{-2}$ is predicted at $z$=2\,kpc. Nevertheless, a somewhat high value $\sim$75 M$_{\sun}$ pc$^{-2}$ is rather predicted at $z=1$\,kpc. For the same reason mentioned previously, our two Kuzmin-Kotozov components modelling of the vertical force is not adequate to correctly reproduce the $K_z$ force in the case of Mondian model that shows  a significant bending of the vertical force law between  $z$=1 and 2\,kpc.

\subsection{A Galactic DM halo with core?}

Now, if we consider the consistency of our local $K_z$ determination with $\Lambda$CDM, we first note that our result is nearly independent from any assumptions on the form of the Galactic potential, for instance in the central regions of the Galaxy. The decomposition of the $K_z$ force in dark and visible contributions requires additional information, for instance, the disc scale length of the visible matter. Previous determinations indicated that there was  less dark matter mass than predicted by cosmological simulations within $R_0$ 
\citep{nav00, bin00,fam05,aba10}, whilst our determination would imply that there is  more dark matter mass inside $R_0$. This has implications for the mass concentration and can be tested relative to the mass-concentration relation reported by \citet{mac08}. This point is dicussed in detail in \citet{pif14b} and they concluded that modifying the Galactic halo profile by taking  NFW adiabatic contraction into account, they can obtain an agreement with simulations in a $\Lambda$CDM universe.
We also remark that the recent Galactic DM halo modelling by \citet{nes13} constrained with inner terminal velocities, MASER observations, and stellar halo velocity dispersions gives a high local DM mass density consistent with our finding. Their modelling favoured a cored profile ($R_H\sim$10\,kpc) of the DM halo
\cite[see also ][]{bis03}.

\section{Conclusion}

We have established that a significant amount of dark matter resides close the Galactic disc: the local dark matter mass density is
$\rho_{\rm DM}(z$=0)= 0.0143$\pm$0.0011M$_{\sun}$\,pc$^{-3}$.
We have  independently determined  the local DM density and the baryonic disc surface density at the solar Galactic radius $R_0$. The large size of our sample leads to small statistical errors, but it is clear that systematic errors could also arise from neglected elements in our modelling. For instance, one aspect of  dynamical modelling can be  questioned: resonant orbits (from vertical relative to  horizontal motions) are numerous at $z$ beyond 1\,kpc and are not modelled by St\"ackel potentials, or with a simple torus fitting. Also, the non-axisymmetric effects due to spiral arms \citep{Faure,Debat} or non-equilibrium features generated by the potential interaction of satellites with the Galactic disc \citep{Gomez} could also bias the result. The  amplitude of non-axisymmetry and non-stationarity and its impact on the $K_z$ studies is in general accepted to be small \citep{rea14}, but should be  quantified precisely on an observational basis. Much larger samples with extremely accurate data on a much wider Galactic volume are expected from the Gaia mission, and will help in examining and solving all such questions.

\begin{acknowledgements}
Funding for RAVE has been provided by the Australian Astronomical	Observatory;	the	Leibniz-Institut	f\"{u}r	Astrophysik Potsdam (AIP); the Australian National University; the Australian Research Council; the French National Research Agency; the German Research Foundation (SPP 1177 and SFB 881); the European Research Council (ERC-StG 240271 Galactica); the Istituto Nazionale di Astrofisica at Padova;  Johns Hopkins University; the National Science Foundation of the USA (AST-0908326); the W. M. Keck foundation; the Macquarie University; the Netherlands Research School for Astronomy; the Natural Sciences and Engineering Research Council of Canada; the Slovenian Research Agency; the Swiss National Science Foundation; the Science \& Technology Facilities Council of the UK; Opticon; Strasbourg Observatory and the universities of Groningen, Heidelberg, and Sydney. The RAVE
website is at http://www.rave-survey.org.

We thank Annie Robin and Michel Cr\'ez\'e for providing figure 11 and for  comments.

\end{acknowledgements}

\bibliographystyle{aa} 
\bibliography{KZ} 

\begin{thebibliography}{56}
\expandafter\ifx\csname natexlab\endcsname\relax\def\natexlab#1{#1}\fi

\bibitem[Abadi et al.(2010)]{aba10} Abadi, M.~G., Navarro, 
J.~F., Fardal, M., Babul, A., \& Steinmetz, M.\ 2010, \mnras, 407, 435 

\bibitem[Batsleer 
\& Dejonghe(1994)]{bat94} Batsleer, P., \& Dejonghe, H.\ 1994, \aap, 287, 43 

\bibitem[Bienaym{\'e}(1999)]{bie99} Bienaym{\'e}, O.\ 1999, \aap, 341, 86 

\bibitem[Bienaym{\'e} et 
al.(2009)]{bie09} Bienaym{\'e}, O., Famaey, B., Wu, X., Zhao, H.~S., \& Aubert, D.\ 2009, \aap, 500, 801 

\bibitem[Bienaym{\'e} et 
al.(2006)]{bie06} Bienaym{\'e}, O., Soubiran, C., Mishenina, T.~V., Kovtyukh, V.~V., \& Siebert, A.\ 2006, \aap, 446, 933 

\bibitem[Binney et al.(2000)]{bin00} Binney, J., Bissantz, 
N., \& Gerhard, O.\ 2000, \apjl, 537, L99 

\bibitem[Binney(2012)]{Binney2012} Binney, J.\ 2012, \mnras, 426, 1324 

\bibitem[Bissantz et al.(2003)]{bis03} Bissantz, N., Englmaier, P., \& Gerhard, O.\ 2003, \mnras, 340, 949 

\bibitem[Bobylev \& Bajkova(2013)]{bob13} Bobylev, V.~V., \& Bajkova, A.~T.\ 2013, Astronomy Letters, 39, 809 

\bibitem[Bovy \& Rix(2013)]{bov13} Bovy, J., \& Rix, H.-W.\ 2013, \apj, 779, 115 

\bibitem[Bovy et al.(2012)]{bov12} Bovy, J., Allende Prieto, C., Beers, T.~C., et al.\ 2012, \apj, 759, 131 

\bibitem[Brunthaler et al.(2011)]{bru11} Brunthaler, A., 
Reid, M.~J., Menten, K.~M., et al.\ 2011, Astronomische Nachrichten, 332, 
461

\bibitem[B{\"u}denbender et al.(2014)]{bud14} B{\"u}denbender, A., van de Ven, G., \& Watkins, L.~L.\ 2014, arXiv:1407.4808 

\bibitem[Burch \& Cowsik(2013)]{bur13} Burch, B., \& Cowsik, R.\ 2013, \apj, 779, 35 

\bibitem[Cabrera-Lavers et al.(2005)]{cab05} Cabrera-Lavers, A., Garz{\'o}n, F., \& Hammersley, P.~L.\ 2005, \aap, 433, 173 

\bibitem[Cannon \& Lloyd(1969)]{can69} Cannon, R.~D., \& Lloyd, C.\ 1969, \mnras, 144, 449 

\bibitem[Chang et al.(2011)]{cha11} Chang, C.-K., Ko, C.-M., \& Peng, T.-H.\ 2011, \apj, 740, 34 

\bibitem[Cr\'ez\'e et al.(1998)]{cre98} Cr\'ez\'e, M., Chereul, E., Bienaym\'e, O., \& Pichon, C.\ 1998, \aap, 329, 920 

\bibitem[Debattista(2014)]{Debat} Debattista, V.~P.\ 2014, arXiv:1405.6345 

\bibitem[Dejonghe 
\& de Zeeuw(1988)]{dej88} Dejonghe, H., \& de Zeeuw, T.\ 1988, \apj, 333, 90 

\bibitem[de Zeeuw(1985)]{dez85} de Zeeuw, T.\ 1985, \mnras, 
216, 273 

\bibitem[ESA(1997)]{esa97} ESA 1997, The Hipparcos catalogue, ESA SP-1200 

\bibitem[Famaey 
\& Dejonghe(2003)]{fam03} Famaey, B., \& Dejonghe, H.\ 2003, \mnras, 340, 752 

\bibitem[Famaey 
\& Binney(2005)]{fam05} Famaey, B., \& Binney, J.\ 2005, \mnras, 363, 603 

\bibitem[Famaey \& McGaugh(2012)]{FamMc12} Famaey, B., \& McGaugh, S.~S.\ 2012, Living Reviews in Relativity, 15, 10 

\bibitem[Faure et al.(2014)]{Faure} Faure, C., Siebert, A., \& Famaey, B.\ 2014, \mnras, 440, 2564 
\bibitem[Flynn et al.(2006)]{fly06} Flynn, C., Holmberg, J.,
  Portinari, L., Fuchs, B., \& Jahrei{\ss}, H.\ 2006, \mnras, 372,
  1149

\bibitem[Foreman-Mackey et al.(2013)]{for13} Foreman-Mackey, D., Hogg, D.~W., Lang, D., \& Goodman, J.\ 2013, \pasp, 125, 306 

\bibitem[Garbari et al.(2012)]{gar12} Garbari, S., Liu, C., 
Read, J.~I., \& Lake, G.\ 2012, \mnras, 425, 1445 

\bibitem[Girardi(1999)]{gir99} Girardi, L.\ 1999, \mnras, 308, 818 

\bibitem[Girardi et 
al.(2000)]{gir00} Girardi, L., Bressan, A., Bertelli, G., \& Chiosi, C.\ 2000, \aaps, 141, 371 

\bibitem[G{\'o}mez et al.(2013)]{Gomez} G{\'o}mez, F.~A., Minchev, I., O'Shea, B.~W., et al.\ 2013, \mnras, 429, 159 

\bibitem[Groenewegen(2008)]{gro08} Groenewegen, M.~A.~T.\ 2008, \aap, 488, 935 

\bibitem[Hammer et al.(2012)]{ham12} Hammer, F., Puech, M., 
Flores, H., et al.\ 2012, European Physical Journal Web of Conferences, 19, 
1004 

\bibitem[Holmberg 
\& Flynn(2004)]{hol04} Holmberg, J., \& Flynn, C.\ 2004, \mnras, 352, 440 

\bibitem[Holmberg et al.(2006)]{hol06} Holmberg, J., Flynn, 
C., \& Portinari, L.\ 2006, \mnras, 367, 449 

\bibitem[James (2004)]{jam04} James, F. \ 2004, 
Reprinted from the Proceedings of the 1972 CERN Computing and Data Processing School

\bibitem[Juri{\'c} et al.(2008)]{jur08} Juri{\'c}, M., Ivezi{\'c}, {\v Z}., Brooks, A., et al.\ 2008, \apj, 673, 864 

\bibitem[Ibata et al.(2013)]{iba13} Ibata, R., Lewis, G.~F., Martin, N.~F., Bellazzini, M., \& Correnti, M.\ 2013, \apjl, 765, L15 

\bibitem[Kapteyn(1922)]{kap22} Kapteyn, J.~C.\ 1922, \apj, 55, 302 

\bibitem[Korchagin et al.(2003)]{kor03} Korchagin, V.~I., 
Girard, T.~M., Borkova, T.~V., Dinescu, D.~I., 
\& van Altena, W.~F.\ 2003, \aj, 126, 2896 

\bibitem[Koposov et al.(2010)]{kop10} Koposov, S.~E., Rix, H.-W., \& Hogg, D.~W.\ 2010, \apj, 712, 260 

\bibitem[Kuijken \& Gilmore(1989)]{kui89} Kuijken, K., \& Gilmore, G.\ 1989, \mnras, 239, 571 

\bibitem[Kuzmin \& Kutuzov(1962)]{kuz62}
Kuzmin, G.~G., \& Kutuzov, S.~A.\ 1962, Bull Abastumani Ap. Obs., 27, 82

\bibitem[Kordopatis et al.(2013)]{kor13} Kordopatis, G., 
Gilmore, G., Steinmetz, M., et al.\ 2013, \aj, 146, 134 

\bibitem[Laney et al.(2012)]{lan12} Laney, C.~D., Joner, 
M.~D., \& Pietrzy{\'n}ski, G.\ 2012, \mnras, 419, 1637 

\bibitem[Lewis 
\& Freeman(1989)]{lew89} Lewis, J.~R., \& Freeman, K.~C.\ 1989, \aj, 97, 139 

\bibitem[Ling(2010)]{lin10} Ling, F.-S.\ 2010, \prd, 82, 023534 

\bibitem[Macci{\`o} et al.(2007)]{mac07} Macci{\`o}, A.~V., Dutton, A.~A., van den Bosch, F.~C., et al.\ 2007, \mnras, 378, 55 

\bibitem[Macci{\`o} et al.(2008)]{mac08} Macci{\`o}, A.~V., 
Dutton, A.~A., \& van den Bosch, F.~C.\ 2008, \mnras, 391, 1940 

\bibitem[McMillan(2011)]{mcm11} McMillan, P.~J.\ 2011, \mnras, 414, 2446 

\bibitem[Miglio et al.(2013)]{mig13} Miglio, A., Chiappini, C., Morel, T., et al.\ 2013, \mnras, 429, 423

\bibitem[Mignard(2000)]{mig00} Mignard, F.\ 2000, \aap, 354, 522 

\bibitem[Milgrom(1983)]{Mil83} Milgrom, M.\ 1983, \apj, 270, 365

\bibitem[Nataf et al.(2012)]{nat12} Nataf, D.~M., Gould, A., \& Pinsonneault, M.~H.\ 2012, \actaa, 62, 33 

\bibitem[Navarro 
\& Steinmetz(2000)]{nav00} Navarro, J.~F., \& Steinmetz, M.\ 2000, \apj, 528, 607 

\bibitem[Nesti \& Salucci(2013)]{nes13} Nesti, F., \& Salucci, P.\ 2013, \jcap, 7, 16 

\bibitem[Ojha(2001)]{ojh01} Ojha, D.~K.\ 2001, \mnras, 322, 
426 

\bibitem[Ojha et 
al.(1996)]{ojh96} Ojha, D.~K., Bienaym\'e, O., Robin, A.~C., Cr\'ez\'e, M., \& Mohan, V.\ 1996, \aap, 311, 456 

\bibitem[Oort(1932)]{oor32} Oort, J.~H.\ 1932, \bain, 6, 249 

\bibitem[Olling \& Dehnen(2003)]{oll03} Olling, R.~P., \& Dehnen, W.\ 2003, \apj, 599, 275 

\bibitem[Olling \& Merrifield(1998)]{oll98} Olling, R.~P., \& Merrifield, M.~R.\ 1998, \mnras, 297, 943 

\bibitem[Pasetto et 
al.(2012a)]{pas12a} Pasetto, S., Grebel, E.~K., Zwitter, T., et al.\ 2012, \aap, 547, A71 

\bibitem[Pasetto et 
al.(2012b)]{pas12b} Pasetto, S., Grebel, E.~K., Zwitter, T., et al.\ 2012, \aap, 547, A70 

\bibitem[Perryman et 
al.(2001)]{per01} Perryman, M.~A.~C., de Boer, K.~S., Gilmore, G., et al.\ 2001, \aap, 369, 339 

\bibitem[Piffl et 
al.(2014a)]{pif14} Piffl, T., Scannapieco, C., Binney, J., et al.\ 2014a, \aap, 562, A91 

\bibitem[Piffl et 
al.(2014b)]{pif14b} Piffl, T., Binney, J., McMillan, P., et al.\ 2014b, arXiv 1404:4130v1  

\bibitem[Pillepich et al.(2014)]{pil14} Pillepich, A., Kuhlen, M., Guedes, J., \& Madau, P.\ 2014, \apj, 784, 161 

\bibitem[Polido et al.(2013)]{pol13} Polido, P., Jablonski, F., \& L{\'e}pine, J.~R.~D.\ 2013, \apj, 778, 32 

\bibitem[Read(2014)]{rea14} Read, J.~I.\ 2014, Journal of Physics G Nuclear Physics, 41, 063101

\bibitem[Read et al.(2009)]{rea09} Read, J.~I., Mayer, L., Brooks, A.~M., Governato, F., \& Lake, G.\ 2009, \mnras, 397, 44 

\bibitem[Reid et al.(2014)]{rei14} Reid, M.~J., Menten, K.~M., Brunthaler, A., et al.\ 2014, \apj, 783, 130 

\bibitem[Robin et al.(2014)]{rob14} Robin, A.C., 2014, in prep

\bibitem[Salucci et al.(2010)]{sal10} Salucci, P., Nesti, F., Gentile, G., \& Frigerio Martins, C.\ 2010, \aap, 523, A83 

\bibitem[Siebert et al.(2003)]{sie03} Siebert, A., Bienaym{\'e}, O., \& Soubiran, C.\ 2003, \aap, 399, 531 


\bibitem[Siebert et al.(2008)]{sie08} Siebert, A., 
Bienaym{\'e}, O., Binney, J., et al.\ 2008, \mnras, 391, 793 

\bibitem[Siebert et al.(2011)]{Sie11a} Siebert, A., Famaey, 
B., Minchev, I., et al.\ 2011, \mnras, 412, 2026 

\bibitem[Siebert et al.(2011)]{sie11} Siebert, A., Williams, 
M.~E.~K., Siviero, A., et al.\ 2011, \aj, 141, 187 

\bibitem[Shu(1969)]{shu69} Shu, F.~H.\ 1969, \apj, 158, 505 

\bibitem[Skrutskie et al.(2006)]{skr06} Skrutskie, M.~F., 
Cutri, R.~M., Stiening, R., et al.\ 2006, \aj, 131, 1163 


\bibitem[Smith et al.(2007)]{smi07} Smith, M.~C., Ruchti, 
G.~R., Helmi, A., et al.\ 2007, \mnras, 379, 755 

\bibitem[Smith et al.(2012)]{smi12} Smith, M.~C., Whiteoak, S.~H., \& Evans, N.~W.\ 2012, \apj, 746, 181 

\bibitem[Soubiran et 
al.(2003)]{sou03} Soubiran, C., Bienaym{\'e}, O., \& Siebert, A.\ 2003, \aap, 398, 141 

\bibitem[Soubiran et 
al.(2008)]{sou08} Soubiran, C., Bienaym{\'e}, O., Mishenina, T.~V., \& Kovtyukh, V.~V.\ 2008, \aap, 480, 91 

\bibitem[Statler(1989)]{sta89} Statler, T.~S.\ 1989, \apj, 
344, 217 

\bibitem[Steinmetz et al.(2006)]{ste06} Steinmetz, M., 
Zwitter, T., Siebert, A., et al.\ 2006, \aj, 132, 1645 

\bibitem[Stello et al.(2013)]{ste13} Stello, D., Huber, D., Bedding, T.~R., et al.\ 2013, \apjl, 765, L41 

\bibitem[Valentini 
\& Munari(2010)]{val10} Valentini, M., \& Munari, U.\ 2010, \aap, 522, A79 

\bibitem[van der Kruit 
\& Freeman(2011)]{kru01} van der Kruit, P.~C., \& Freeman, K.~C.\ 2011, \araa, 49, 301 

\bibitem[van Leeuwen(2007)]{van07} van Leeuwen, F.\ 2007, \aap, 474, 653 

\bibitem[Veltz et 
al.(2008)]{vel08} Veltz, L., Bienaym{\'e}, O., Freeman, K.~C., et al.\ 2008, \aap, 480, 753 

\bibitem[Williams et al.(2013)]{wil13} Williams, M.~E.~K., 
Steinmetz, M., Binney, J., et al.\ 2013, \mnras, 436, 101 

\bibitem[Zhang et al.(2013)]{zha13} Zhang, L., Rix, H.-W., 
van de Ven, G., et al.\ 2013, \apj, 772, 108 

\bibitem[Zhao et al.(2000)]{zha00} Zhao, G., Qiu, H.-M., \& Zhang, H.-W.\ 2000, Acta Astrophysica Sinica, 20, 389

\bibitem[Zwitter et al.(2008)]{zwi08} Zwitter, T., Siebert, 
A., Munari, U., et al.\ 2008, \aj, 136, 421 





\end{thebibliography}

\appendix
 
 \section{Separable potential and Jeans equation of the vertical motion}
 
 (Summary)
 Here we show that the Jeans equation of the vertical motion of stars and the Collisionless Boltzmann equation (CBE) for the vertical motions have the same general solutions. The CBE is  obtained under the assumption of separability of the vertical and horizontal motions (i.e. $\sigma_{r,z}=0$), while the Jeans equation is obtained under a less restrictive assumption.
 
 \citet{gar12} used a  general formulation for the solutions of the Jeans equation of vertical motion. Then, they claim that their solution is more general than that obtained under the more restrictive hypothesis of separability.

Here, we show the opposite. Under these two different hypotheses (one  more restrictive than the other), we obtain the same general solutions. Thus, the  useful formulation used by \citet{gar12}  is not more general than that obtained under the simple hypothesis of separability. \\

 (Justification)
 The usual  approximation, at low $z$, to model the motion of stars perpendicular to the Galactic plane consists of assuming that the Galactic potential is separable in $R$ and $z$ coordinates, thus the correlation between vertical and horizontal velocities is zero, $\sigma_{Rz}=0$ and the velocity ellipsoids remain parallel to the Galactic plane. 
  Reciprocally, the  potential is separable in the domains where $\sigma_{Rz}=0$. 
 
 Under the assumption of separability, the stationary vertical distribution function of stars may be described using the 2D collisionless Boltzmann equation (CBE):
 
 \begin{equation}
 w\,\frac{\partial f(z,w)}{\partial z}-\frac{\partial \Phi(z)}{\partial z} \,\frac{\partial f(z,w)}{\partial w}=0\, .
 \label{eqB1}
  \end{equation}
  
Besides, the  Jeans equation corresponding to  stationary vertical motions is
 
 \begin{equation}
 \frac{1}{R}\, \frac{\partial }{\partial R}\left( R\nu\sigma_{Rz}^2 \right)
 + \frac{\partial }{\partial z}\left( \nu\sigma_{z}^2 \right)
 + \nu \frac{\partial \Phi}{\partial z} = 0 
 \label{eqB2}
   \end{equation}
   
 and can also be simplified by canceling the first l.h.s. term
  when the potential is separable since then $\sigma_{Rz}=0$.
 
 Recently, \citet{gar12}  used a    general solution of this reduced  Jeans equation:
\begin{equation}
  \frac{\partial }{\partial z}\left( \nu\sigma_{z}^2 \right)
 + \nu \frac{\partial \Phi}{\partial z} = 0\, .
  \label{eqB3}
   \end{equation}
  
 They notice that the cancellation of the first l.h.s. term of Equation \ref{eqB2} covers more  general and less  restricting hypothesis than the assumption of separability (i.e. $\sigma_{Rz}=0$) 
 and claim that this " minimal assumption method ... breaks the assumption that the distribution is separable".
  Without a clear justification,  they assume that their  general solution is more general than those usually  used, for instance by \citet{hol04}.
 
 Here, we show the opposite and we establish that  cancelling the l.h.s. term of the Jeans Equation \ref{eqB1} gives the same solution that
 in the case of separability of the $R$ and $z$ motions. \\
  
 Thus, let be $f(z$=0,$w$) = $g^*(p$=$w^2/2)$, the velocity distribution of a stationary solution  at $z$=0.
  This function is odd and can be  written as an integration over an infinite set  of Gaussians:

\begin{equation}
f(z=0,w)=g^*(p)=\int_0^{\infty} a(\sigma) \,\mathrm{e}^{-w^2/(2\sigma^2)} \,d\sigma 
 \label{eqB4}\,.
   \end{equation}
  
This is equivalent to the Laplace transform, with $\beta=1/\sigma^2$ and $a(\sigma)\,d\sigma=g(\beta)d\beta$:

\begin{equation}
g^*(p)=\int_0^{\infty} g(\beta) \,\mathrm{e}^{-\beta p} \,d\beta = \mathcal{L} \,[g(\beta)]\,. 
 \label{eqB5}
   \end{equation}

If the integral exists, for instance when $f(z=0,w)$ is null for large values of $w$, then $g(\beta)$ exists,  it is unique and is given by the inverse Laplace transform

\begin{equation}
 g(\beta) = \mathcal{L}^{-1}[g^*(p)]
 \label{eqB6}
    \end{equation}
  that gives us the unique decomposition in Gaussians (Equation \ref{eqB4}).

For an isothermal  component (i.e. Gaussian in velocities with a dispersion $\sigma$) the solution of the Jeans Equation \ref{eqB3} is  $\nu_{\sigma}(z)=\rm{e}^{-\beta \Phi(z)}$, with $\Phi$(0)=0,
and since, from Equation \ref{eqB4}, we have
$$\nu(0)=\int_0^{\infty}  a(\sigma)\,\rm{d} \sigma $$
 then
the general solution  can  be written:

\begin{equation}
 \nu(z)=\int_0^{\infty} a(\sigma) \, \mathrm{e}^{-\Phi(z)/\sigma^2} \, d\sigma \,.
 \label{eqB7}
  \end{equation}

This  general solution of Equation \ref{eqB3} has a different form, but is identical to the general solution given by the equation 8 of \citet{gar12}. A key point is that for any given odd function $f(z$=0,$w)$ representing the distribution function at $z$=0,
there is a unique  function $f'(z,w)$, solution of the Jeans Equation \ref{eqB3}, that satisfies $f'(z$=0,$w$) = $f(z$=$0,w)$.\\

Now, on the other hand,  the general solution of the CBE  is  $h(E)$ where $E$ is the energy.  The distribution function  at $z$=0,
is given by $h(E(z$=0)) = $h(w^2/2)$. As  has been done previously, we can inverse the function $h$   and rewrite the general solution.
Thus, we
 obtain an equivalent form of the general solution:

\begin{equation}
 f(z,w)=\int_0^{\infty} a(\sigma) \, \mathrm{e}^{-\Phi(z)/\sigma^2} \, \frac{1}{\sqrt{2 \pi} \sigma}\mathrm{e}^{-w^2/(2\sigma^2)} d\sigma\, , 
 \label{eqB8}
   \end{equation}
  
where $a(\sigma)$ is related through an inverse Laplace transform to  $h(E$=0).

By integrating $f(z,w)$ over the $w$ velocities, we exactly recover the general solution (Equation \ref{eqB7}) of the Jeans Equation \ref{eqB3}.

Thus, the general solution obtained for the  Jeans Equation \ref{eqB3} has the same form as the general solution of the CBE for a separable potential.\\

In conclusion,  the hypothesis that the l.h.s. term in Equation \ref{eqB2}  is null,  is indeed a more general hypothesis than the hypothesis of separability. However, contrary to the  \citet{gar12} claim, the general solution of Equation \ref{eqB3}   is   neither more general nor different than the solution (Equation \ref{eqB8}) obtained from  the hypothesis of separability.

\end{document}